\documentclass[aps,prd,preprintnumbers,nofootinbib,superscriptaddress,
fleqn,floatfix,tightenlines,twocolumn,compress,10pt]{revtex4} 
\usepackage{amsmath,amsfonts,amssymb,amscd,amsxtra,amsthm}
\usepackage{graphicx}  
\usepackage{epstopdf}
\usepackage{dcolumn}  
\usepackage{bm}          
\usepackage{slashed}
\usepackage{cancel}
\usepackage{float}
\usepackage{mathtools}
\usepackage{amsbsy}
\usepackage{amstext}
\usepackage{tabularx}
\usepackage{enumitem}  
\usepackage{array} 
\usepackage{multirow}
\usepackage{physics}
\usepackage{makecell}
\usepackage{subcaption}
\usepackage{dsfont}
\usepackage{hyperref}

\usepackage[utf8]{inputenc} 
\usepackage{booktabs} 
\usepackage[normalem]{ulem} 
\usepackage{xcolor} 

\renewcommand{\sout}{\bgroup \color{red} \ULdepth=-.5ex \ULset}

\makeatletter

\begin{document}
\title{Transverse energy-momentum tensor distributions in polarized nucleons} 
\author{Ho-Yeon Won}
\email[E-mail: ]{hoyeon.won@polytechnique.edu}
\affiliation{CPHT, CNRS, \'Ecole polytechnique, Institut Polytechnique de Paris,
Palaiseau, France}
 
\author{C\'edric Lorc\'e}
\email[E-mail: ]{cedric.lorce@polytechnique.edu}
\affiliation{CPHT, CNRS, \'Ecole polytechnique, Institut Polytechnique de Paris,
Palaiseau, France}
\date {\today}
\begin{abstract}
We complete our study of the relativistic spatial distributions of the energy-momentum tensor inside polarized nucleons within the quantum phase-space formalism. In the present work, we focus on the components of the energy-momentum tensor involving at least one transverse index.
We also explore the multipole structure of the transverse distributions in a moving nucleon.
In the infinite-momentum frame, we show that the formalism reproduces the standard light-front distributions, including those with a ``bad'' component, and explains the origin of their structure.
\end{abstract}
\maketitle
\section{Introduction}
One of the most prominent challenges in hadron physics is understanding the internal structure of nucleons.
In particular, the energy-momentum tensor (EMT) provides a unified framework for exploring mass~\cite{Ji:1995sv,Lorce:2017xzd,Hatta:2018sqd,Metz:2020vxd,Lorce:2021xku,Liu:2021gco}, angular momentum~\cite{Jaffe:1989jz,Ji:1996ek,Shore:1999be,Bakker:2004ib,Lorce:2018zpf,Ji:2020hii,Lorce:2021gxs}, and mechanical properties~\cite{Polyakov:2002yz,Polyakov:2018zvc} inside the nucleon, see also the reviews~\cite{Leader:2013jra,Burkert:2023wzr,Lorce:2025aqp}.
The spatial distributions of the EMT have been extensively studied from theoretical perspectives in, e.g., 
Refs.~\cite{Abidin:2008sb,Chakrabarti:2015lba,Lorce:2017wkb,Lorce:2018egm,Schweitzer:2019kkd,Cosyn:2019aio,Kim:2020lrs,Chakrabarti:2020kdc,Panteleeva:2021iip,Freese:2021czn,Freese:2021mzg,Pefkou:2021fni,More:2021stk,Choudhary:2022den,Won:2022cyy,More:2023pcy,Won:2023zmf,Lorce:2025oot,Won:2025dgc,Freese:2025tqd,Lorce:2025pxt,Kim:2025iis,Tanaka:2025pny,Sain:2025kup,Fujii:2025eug,Fujii:2025pkv,Fujii:2025aip,Fukushima:2025jah,Stegeman:2025sca,Stegeman:2025tdl,Fukushima:2026wwc}, and their precise mapping constitutes one of main goals of the upcoming Electron-Ion Collider (EIC)~\cite{Accardi:2012qut,Aschenauer:2017jsk,AbdulKhalek:2021gbh}.
Furthermore, the parity-odd partner of the quark EMT has recently been proposed~\cite{Lorce:2014mxa} as a tool to investigate spin-orbit correlations, and more generally the chiral structure of hadrons~\cite{Kim:2024cbq,Lorce:2025ayr}.

Spatial distributions in quantum mechanics are typically defined via Fourier transforms of form factors (FFs).
They were first used to study the three-dimensional electromagnetic structure of the nucleon in the Breit frame (BF)~\cite{Breit:1934zz,Ernst:1960zza,Sachs:1962zzc}, where the nucleon is in average at rest.
The definition is strictly valid only in the non-relativistic regime, when the particle size is much larger than its reduced Compton wavelength.
In particular, it does not apply to the nucleon, whose charge radius turns out to be comparable to its reduced Compton wavelength~\cite{Xiong:2019umf} (see also Ref.~\cite{ParticleDataGroup:2024cfk} for more discussions and results). This implies that relativistic recoil corrections become non-negligible in the momentum transfer regime $Q \gtrsim 2M$, calling for a proper treatment within quantum field theory~\cite{Yennie:1957skg,Kelly:2002if,Miller:2018ybm,Jaffe:2020ebz}.
As a result, an alternative definition of BF distributions was proposed in Refs.~\cite{Friar1975,Lorce:2018egm,Lorce:2020onh}.

The quantum phase-space formalism allows one to extend the concept of relativistic spatial distribution to a larger class of frames, where the nucleon may in general have a non-vanishing average momentum. However, to preserve a time-independent picture, these distributions must be integrated over the longitudinal coordinate. Since the first application to the study of longitudinal angular momentum (AM)~\cite{Lorce:2017wkb}, this formalism has enabled the construction of relativistic spatial distributions in the transverse plane for various operators, including the electromagnetic current~\cite{Lorce:2020onh,Kim:2021kum,Lorce:2022jyi,Chen:2022smg,Kim:2022wkc,Chen:2023dxp,Hong:2023tkv}, the axial-vector current~\cite{Chen:2024oxx,Chen:2024ksq}, and the EMT~\cite{Lorce:2018egm,Kim:2022wkc,Won:2022cyy,Won:2025dgc}.

In Ref.~\cite{Lorce:2018egm}, the EMT study focused on unpolarized nucleons. We recently extended in Ref.~\cite{Won:2025dgc} the analysis to polarized nucleons, but we considered only the subset of EMT components that do not involve any transverse index. In the present work, we complete our study and discuss the other EMT components. This paper is organized as follows.
In Sec.~\ref{sec.2}, we briefly review the asymmetric, local and gauge invariant EMT operator, the parametrization of its matrix
elements, and some of its key properties. 
Sec.~\ref{sec.3} provides the definition of the relativistic spatial distributions within the quantum phase-space approach.
There, we analyze the EMT components with at least one transverse index, considering both longitudinal and transverse nucleon polarizations.
In Sec.~\ref{sec.4}, we present the light-front (LF) amplitudes and compare them with previous results in the infinite-momentum frame (IMF). Finally, we summarize our results in Sec.~\ref{sec.5}.
In the present work, we use the conventions $\epsilon_{0123} = \epsilon^{123}=1$ 
and the notations $a^{[\mu}b^{\nu]}\equiv a^{\mu}b^{\nu}-a^{\nu}b^{\mu}$ and $a^{\{\mu}b^{\nu\}}\equiv a^{\mu}b^{\nu}+a^{\nu}b^{\mu}$.

\section{Energy-momentum tensor operators and its matrix elements \label{sec.2}}
In the present work, we adopt the asymmetric and gauge-invariant form of the local EMT current~\cite{Bakker:2004ib,Lorce:2017wkb,Lorce:2018egm} (see the review~\cite{Leader:2013jra} for a discussion of other EMT forms).
In quantum chromodynamics (QCD), the quark and gluon parts of the EMT current are defined by 
\begin{align}
    T_{q}^{\mu\nu}(x)
& =   \bar{\psi}(x) 
    \gamma^{\mu}    
    \frac{i}{2}  \overleftrightarrow{D}^{\nu} 
    \psi(x), \\
    T_{G}^{\mu\nu}(x)
& = 
    F^{\mu\lambda,a}(x)  F_{\lambda}^{\phantom{a}\nu,a}(x) \nonumber \\
  &+ \frac{1}{4}\,
    g^{\mu\nu}
    F^{\lambda\rho,a}(x)  F_{\lambda\rho}^{a}(x).
\end{align}
where $\overleftrightarrow{D}_{\mu}=\overrightarrow{\partial}_{\mu}-\overleftarrow{\partial}_{\mu}-2igA_{\mu}^{a}T^{a}$ denotes the two-sided covariant derivative.                         
For simplicity, we will omit in the subscript $q,G$ and specify it only when needed.
The gluon field is expressed as $A^{\mu}=A^{\mu,a}T^{a}$ with respect to the SU(3) generators $T^{a}$ in the adjoint representation.
The corresponding field-strength tensor is $F_{\mu\nu}^{a}=\partial_{\mu}A_{\nu}^{a}-\partial_{\nu}A_{\mu}^{a}+gf^{abc}A_{\mu}^{b}A_{\nu}^{c}$, where $f^{abc}$ are the fully antisymmetric SU(3) structure constant.

The total (i.e., quark + gluon) EMT satisfies the conservation law
\begin{align}
    \partial_{\mu} T^{\mu\nu}=0,
\label{EMT_conservation}
\end{align}
and consequently four-momentum $P^{\mu}
  = \int d^{3}x\,
    T^{0\mu} (x)$ is conserved. 
Also, the antisymmetric part of the quark EMT is related to the axial-vector current $J_{5}^{\mu}=\bar{\psi}\gamma^{\mu}  \gamma_{5} \psi$ via QCD equations of motion~\cite{Shore:1999be,Bakker:2004ib,Leader:2013jra,Lorce:2017wkb}
\begin{align}
    T_{q}^{ [\mu \nu ] } 
    (x)
& = - \frac{1}{2} \epsilon^{\mu \nu \alpha \beta}     \partial_{\alpha} 
    J_{5,\beta} 
    (x).
\label{anti_currents}
\end{align}

The matrix elements of the asymmetric EMT for a spin-$\frac{1}{2}$ state with mass $M$ can be parameterized in terms of multipole FFs, similar to those for the EM current~\cite{Lorce:2020onh}:
\begin{widetext}
\begin{align}
    \mel{p^{\prime},s^{\prime}}{T^{\mu\nu}(0)}{p,s} 
& = \bar{u}\left(p^{\prime},s^{\prime}\right)
    \Bigg[
    M\,\frac{P^{\mu}P^{\nu}}{P^2}
    E(Q^2)  
  + \frac{\Delta^{\mu}\Delta^{\nu}-\frac{1}{2}g^{\mu\nu}_P\Delta^{2}}{4M}  
    F_{2}(Q^2)
  - M g^{\mu\nu}_P 
    F_{0}(Q^2)\notag\\
&   \hspace{2.2cm}
  + \frac{iP^{\{\mu}\epsilon^{\nu\}\alpha\beta\rho}\Delta_{\alpha}P_\beta\gamma_\rho\gamma_5}{2P^2} J(Q^2)
  - \frac{i}{2}\epsilon^{\mu\nu\alpha\beta}\Delta_\alpha\gamma_\beta\gamma_5\,
    S(Q^2)
    \Bigg]
    u\left(p,s\right).
\label{Parametrization2}
\end{align}
\end{widetext}
The momentum states are normalized as
$\braket{p^{\prime},s^{\prime}}{p,s}=
2P^{0}\left(2\pi\right)^{3}\delta^{(3)}
(\bm{p}^{\prime}-\bm{p})\,\delta_{s^{\prime}s}$,
and the Dirac spinors are normalized as $\bar{u}\left(p,s^{\prime}\right)u\left(p,s\right)=2M\delta_{s^{\prime}s}$ with $s^{\prime},s$ denoting the canonical polarizations.
The average momentum and momentum transfer are defined by $P=\left(p^{\prime}+p\right)/2$ and $\Delta=p^{\prime}-p$,
respectively.
$g^{\mu\nu}_P=g^{\mu\nu}-P^\mu P^\nu/P^2$ 
is the projector onto the subspace orthogonal to $P$. 

In the transverse BF defined by the conditions $\Delta_{z}=0$ and $\bm{P}=\bm{0}$~\cite{Won:2025dgc}, the EMT matrix elements $\langle\langle T^{\mu\nu}\rangle\rangle=\mel{p^{\prime},s^{\prime}}{T^{\mu\nu}(0)}{p,s}/(2P^{0})$ read~\cite{Polyakov:2018zvc} 
\begin{subequations}\label{BFampl}
\begin{align}
    \hspace{-0.5cm}
    \langle\langle
    T^{00}
    \rangle\rangle_{\mathrm{BF}} 
& = M\,
    \delta_{s^{\prime}s}\,
    E \,, \\
    \hspace{-0.5cm}
    \langle\langle
    T^{0i}
    \rangle\rangle_{\mathrm{BF}} 
& = -M\,
    \sigma_{s^{\prime}s}^{3} \,  
   i\epsilon^{ij}_\perp X_{1}^{j}(\phi_{\bm{\Delta}})\,
     \sqrt{\tau} \left(J -S \right), \\
         \hspace{-0.5cm}
    \langle\langle
    T^{03}
    \rangle\rangle_{\mathrm{BF}} 
& = M\,
    \sigma_{s^{\prime}s}^{i} \,  
   i\epsilon^{ij}_\perp X_{1}^{j}(\phi_{\bm{\Delta}})\,
     \sqrt{\tau} \left(J -S \right), \\
    \hspace{-0.5cm}
    \langle\langle
    T^{i0}
    \rangle\rangle_{\mathrm{BF}} 
& = -M\,
    \sigma_{s^{\prime}s}^{3} \,  
   i\epsilon^{ij}_\perp X_{1}^{j}(\phi_{\bm{\Delta}})\,
     \sqrt{\tau} \left(J +S \right), \\
         \hspace{-0.5cm}
    \langle\langle
    T^{30}
    \rangle\rangle_{\mathrm{BF}} 
& = M\,
    \sigma_{s^{\prime}s}^{i} \,  
   i\epsilon^{ij}_\perp X_{1}^{j}(\phi_{\bm{\Delta}})\,
     \sqrt{\tau} \left(J +S \right), \\     
    \hspace{-0.5cm}
    \langle\langle
    T^{ij}
    \rangle\rangle_{\mathrm{BF}} 
& = M\,
    \delta_{s^{\prime}s}
    \left[
    \delta_{\perp}^{ij} \,
    F_{0} 
  + X_{2}^{ij}(\phi_{\bm{\Delta}})\,
   \tau  F_{2} 
    \right], \\
    \hspace{-0.5cm}
    \langle\langle
    T^{33}
    \rangle\rangle_{\mathrm{BF}} 
& = M\,
    \delta_{s^{\prime}s}
    \left(
    F_{0} 
  - \frac{\tau}{2} 
    F_{2} 
    \right), \\
  \hspace{-0.5cm} 
    \langle\langle
    T^{3i}
    \rangle\rangle_{\mathrm{BF}} 
& = \langle\langle
    T^{i3}
    \rangle\rangle_{\mathrm{BF}}=0,
\end{align}
\end{subequations}
where we used for convenience the Pauli matrix elements $\bm{\sigma}_{s^\prime s}$ and the dimensionless, Lorentz-invariant quantity $\tau=Q^{2}/(4M^{2})$. Latin indices $i,j,\cdots$ denote in this work transverse components.
We also introduced irreducible symmetric-traceless tensors of rank $n$ in the transverse plane~\cite{Kim:2022bia,Kim:2022wkc,Hong:2023tkv} to reveal the multipole structure. For instance, 
\begin{subequations}
\begin{align}
    \hspace{-0.5cm}
    X_{1}^{i}
    (\phi_{\bm{v}})
& = \frac{v_{\perp}^{i}}{\left|\bm{v}_{\perp}\right|}=\left(\cos\phi_{\bm{v}},\sin\phi_{\bm{v}}\right),\\
    \hspace{-0.5cm}
    X_{2}^{ij}
    (\phi_{\bm{v}})
& = \frac{v^i_\perp v^j_\perp}{|\bm{v}_\perp|^2}
  - \frac{1}{2}\,
    \delta_{\perp}^{ij}, \\
    \hspace{-0.5cm}
    X_{3}^{ijk}
    (\phi_{\bm{v}})
& = \frac{v^i_\perp v^j_\perp v^k_\perp}{\left|\bm{v}_{\perp}\right|^3} 
  - \frac{1}{4} \frac{\delta_{\perp}^{ij} v^k_\perp
  + \delta_{\perp}^{jk} v^i_\perp
  + \delta_{\perp}^{ki} v^j_\perp}{\left|\bm{v}_{\perp}\right|}  .
\end{align}
\end{subequations}
The multipole FFs are linear combinations of the standard EMT FFs~\cite{Ji:1996ek,Polyakov:2002yz}
\begin{subequations}
\begin{align}
    E
& = A
  + \bar{C}
  + \tau
    \left( 
    A
  - 2 J
  + D
    \right), \\
    J
& = (A+B)/2, \\
    F_{0}
& = - \bar{C}
  - \tau D/2, 
    \hspace{1cm}
    F_{2}
  = D.
\end{align}
\end{subequations}

It follows from Poincar\'e symmetry that the total multipole FFs must satisfy the constraints~\cite{Ji:1996ek,Leader:2013jra,Teryaev:1999su,Lowdon:2017idv,Cotogno:2019xcl}
\begin{equation}
\begin{aligned}
  \hspace{1.5cm}&  E
    (0)
  = 1,  
    \qquad
    J
    (0)
  = \frac{1}{2},
\\
   & F_{0}
    (Q^2)+\frac{\tau}{2}F_2(Q^2)
  = 0.
\label{symmetry}
\end{aligned}
\end{equation}
Although the value of $F_{2}(0)$ remains unconstrained by the symmetry, it is conjectured to be negative in QCD based on stability arguments~\cite{Burkert:2023wzr,Perevalova:2016dln,Polyakov:2018zvc,Lorce:2025oot}. 
As a result of Eq.~\eqref{anti_currents}, the quark intrinsic spin FF $S_{q}$ is related to the axial-vector FF as follows~\cite{Shore:1999be,Bakker:2004ib,Leader:2013jra,Lorce:2017wkb}:
\begin{equation}
  S_{q}(Q^{2})
 = \frac{1}{2}\,
  G_{A}^{q}(Q^{2}).
\end{equation}
On the other hand, since there is no gauge-invariant local operator representing the gluon intrinsic spin, we have $S_{G}(Q^{2})
  = 0$~\cite{Leader:2013jra}.

\section{Distributions in elastic frame \label{sec.3}}

The transverse BF belongs to the class of elastic frames (EF), characterized by $\Delta_{z} = 0$ and $\bm{P}=P_z\,\bm{\hat{e}}_{z}$~\cite{Lorce:2017wkb}. Within the quantum phase-space formalism developed in~\cite{Lorce:2018egm}, Fourier transforms of EF matrix elements are interpreted as relativistic spatial distributions in the transverse plane. Due to Lorentz symmetry, these distributions usually do not allow for a genuine (probabilistic) density interpretation, unless one works in the IMF $P_z\to \infty$. The quantum phase-space formalism has been used to investigate the frame dependence of the relativistic spatial distributions of the electromagnetic current~\cite{Lorce:2020onh,Kim:2021kum,Lorce:2022jyi,Chen:2022smg,Kim:2022wkc,Chen:2023dxp,Hong:2023tkv}, axial-vector current~\cite{Chen:2024oxx,Chen:2024ksq}, and EMT~\cite{Lorce:2018egm,Kim:2022wkc,Won:2022cyy,Won:2025dgc}, providing in particular an interpolation between the BF and IMF pictures.

In~\cite{Lorce:2018egm}, the discussion was limited to the case of an unpolarized nucleon and focused on the limits $P_z\to 0$ and $P_z\to \infty$. We recently extended the analysis to the case of a polarized nucleon for any value of $P_z$~\cite{Won:2025dgc}, but we investigated only the subset $\{T^{00},T^{03},T^{30},T^{33}\}$ of EMT components. In the present work, we complete our study of the relativistic spatial distributions of the EMT inside a polarized nucleon, by analyzing the remaining components of the EMT that involve at least one transverse index.

\subsection{Definition of relativistic spatial distributions}
In the quantum phase-space formalism, two-dimensional (2D) relativistic spatial distribution of the EMT are defined as~\cite{Lorce:2017wkb,Lorce:2018egm}
\begin{align}
&   \mathcal T^{\mu\nu}
    (\bm{b}_{\perp},P_{z};s^{\prime},s)\notag\\
& = \int \frac{d^{2}\Delta_{\perp}}{\left(2\pi\right)^{2}}\,
    e^{-i\bm{\Delta}_{\perp}\cdot\bm{b}_{\perp}}
    \langle\langle
    T^{\mu\nu}
    \rangle\rangle_{\mathrm{EF}}.
\label{EMTdistribution_EF}
\end{align}
The impact-parameter coordinates are denoted as $\bm{b}_{\perp}$, and correspond to the position relative to the canonical center of the system~\cite{Jaffe:1989jz,Shore:1999be,Bakker:2004ib,Leader:2013jra,Lorce:2018zpf}.
In the EF, the kinematical variables are given by $P^\mu=\left(P^{0},\bm{0}_{\perp},P_{z}\right)$ and $\Delta^\mu=\left(0,\bm{\Delta}_{\perp},0\right)$.

Under a Lorentz boost satisfying $p^\mu=\Lambda^\mu_{\phantom{\mu}\nu}p^\nu_{\mathrm{BF}}$, the BF matrix elements transform according to~\cite{Jacob:1959at,Durand:1962zza} 
\begin{align}
&   \mel{p^{\prime},s^{\prime}}{T^{\mu\nu}(0)}{p,s}  \notag\\
& = \sum_{s_{\mathrm{BF}}^{\prime},s_{\mathrm{BF}}}
    D_{s_{\mathrm{BF}}s}^{(j)}
    (p_{\mathrm{BF}},\Lambda)\,
    D_{s_{\mathrm{BF}}^{\prime}s^{\prime}}^{*(j)}    
    (p_{\mathrm{BF}}^{\prime},\Lambda) \notag\\
&   \hspace{.5cm}  \times    
    \Lambda_{\phantom{\mu}\alpha}^{\mu}
    \Lambda_{\phantom{\nu}\beta}^{\nu}
    \mel{p_{\mathrm{BF}}^{\prime},s_{\mathrm{BF}}^{\prime}}
    {T^{\alpha\beta}(0)}
    {p_{\mathrm{BF}},s_{\mathrm{BF}}},
\label{LT}
\end{align}
where $D^{(j)}$ is the Wigner rotation matrix for spin-$j$ targets. For spin-1/2 targets, the Wigner rotation matrix takes the form:
\begin{align}
    D_{s^{\prime}s}^{(1/2)}
    (p,\Lambda)
  = \cos{\frac{\theta}{2}}\, \delta_{s^{\prime}s}
  + i \sin{\frac{\theta}{2}}\,
    \frac{\left(\bm{p}\times\bm{\sigma}_{s^{\prime}s}\right)_{z}}{\left|\bm{p}_{\perp}\right|}.
\end{align}
The Wigner rotation angle can be determined from Eq.~\eqref{LT} by matching a direct evaluation of the left-hand side in the EF using Dirac bilinears~\cite{Lorce:2017isp} with the right-hand side obtained from the transverse BF amplitudes in Eq.~\eqref{BFampl}. It depends neither on the specific operator nor on the spin of the target. One consistently finds~\cite{Lorce:2022jyi,Chen:2022smg,Chen:2024oxx,Won:2025dgc}
\begin{align}
    \cos{\theta}
& = \frac{P^{0}+M\left(1+\tau\right)}{\left(P^{0}+M\right)\sqrt{1+\tau}}, \notag\\
    \sin{\theta}
& = - \frac{\sqrt{\tau}P_{z}}{\left(P^{0}+M\right)\sqrt{1+\tau}}.
\label{WR}
\end{align}
The Lorentz boost parameters are given by
\begin{align}
    \gamma
  = \frac{P^{0}}{\sqrt{P^{2}}},
    \qquad
    \beta
  = \frac{P_{z}}{P^{0}},
\label{WR2}
\end{align}
with $P^{2}=M^{2}(1+\tau)$ and $P^{0}=\sqrt{P_{z}^{2}+M^{2}\left(1+\tau\right)}$.
In the forward limit, the Lorentz boost parameters will be denoted by $\gamma_{P}=\gamma|_{\Delta\to0}$ and $\beta_{P}=\beta|_{\Delta\to0}$.

The EMT provides, for instance, the relativistic energy (or inertia) of the system.
The latter becomes, however, infinitely large in the IMF.
Following Refs.~\cite{Lorce:2018egm,Won:2025dgc}, we define normalized EMT distributions by factoring out some Lorentz boost factors from the relativistic EMT distributions in Eq.~\eqref{EMTdistribution_EF}
\begin{align}
    \mathcal T^{\mu\nu}
 := \begingroup
    \renewcommand{\arraystretch}{2} 
    \setlength{\tabcolsep}{15pt} 
    \begin{pmatrix}
    \gamma_{P}\rho            & \mathcal{P}^{i}_\perp & \gamma_{P}\mathcal{P}^{z}\\
    \mathcal{I}^{i}_\perp     & \gamma_{P}^{-1}\Pi^{ij}_\perp    & \Pi^{iz}_\perp \\
    \gamma_{P}\mathcal{I}^{z} & \Pi^{zi}_\perp        & \gamma_{P}\Pi^{zz} 
    \end{pmatrix}
    \endgroup.
\label{defEMTdis}
\end{align}
We classify these normalized EMT distributions according to their transformation properties under rotations about the $z$-axis: scalar ($\rho,\mathcal{P}^{z},\mathcal{I}^{z},\Pi^{zz}$), vector ($\mathcal{P}_{\perp}^{i},\mathcal{I}_{\perp}^{i},\Pi_{\perp}^{iz},\Pi_{\perp}^{zi}$), and tensor ($\Pi_{\perp}^{ij}$).
The scalar distributions were studied in our previous work~\cite{Won:2025dgc}, since the matrix elements of $T^{00},T^{03},T^{30}$, and $T^{33}$ form a closed set under longitudinal Lorentz boosts.
In the present work, we study the vector and tensor distributions.


\subsection{Vector distributions $\mathcal{P}^{i}_\perp,\mathcal{I}^{i}_\perp,\Pi^{zi}_\perp,\Pi^{iz}_\perp$}

We start with the EF matrix elements of $T^{0i}$, $T^{i0}$, $T^{3i}$, and $T^{i3}$. Just like the transverse electromagnetic current~\cite{Kim:2021kum,Chen:2022smg}, they  are purely dipolar. More precisely, we find
\begin{subequations}
\begin{align}
    \langle\langle T^{0i}\rangle\rangle_{\mathrm{EF}}
& = \langle\langle T^{0i}\rangle\rangle_{\mathrm{BF}}, 
    \label{T0i}\\
    \langle\langle T^{3i}\rangle\rangle_{\mathrm{EF}}
& = \beta 
    \langle\langle T^{0i}\rangle\rangle_{\mathrm{BF}},
\label{T3i}
\end{align}
\end{subequations}
and similar relations for $\langle\langle T^{i0}\rangle\rangle_{\mathrm{EF}}$ and $\langle\langle T^{i3}\rangle\rangle_{\mathrm{EF}}$. These matrix elements are proportional to the Pauli matrix elements $\sigma_{s^{\prime}s}^{3}$ and therefore remain invariant under the Wigner rotation, as noted in~\cite{Chen:2022smg}.

The 2D spatial distribution of transverse momentum reads
\begin{equation}
    \mathcal P^i_\perp (\bm{b}_{\perp};s^{\prime},s) =- \sigma_{s^{\prime}s}^{3}\,\epsilon^{ij} _\perp
    X_{1}^{j}(\phi_{\bm{b}})
    \mathcal{P}_{1}
    (b),
\end{equation}
where the (dipolar) radial distribution is given by
\begin{equation}
    \mathcal{P}_{1}
    (b)
  = -\frac{1}{2} \frac{d}{db}
    \int \frac{d^{2}\Delta_{\perp}}{\left(2\pi\right)^{2}} \,
    e^{-i\bm{\Delta}_{\perp}\cdot\bm{b}_{\perp}} 
    \,\tilde{\mathcal{P}}_{1}
    (Q^{2}),
\end{equation}
with $\tilde{\mathcal{P}}_{1}
    (Q^{2})
=J(Q^{2})
  - S(Q^{2})$. Remarkably, $\bm{\mathcal P}_\perp$ does not depend on $P_z$. While the total transverse momentum vanishes by definition of the EF,
\begin{equation}
    \bm{P}_\perp:=\int d^2b_\perp\,\bm{\mathcal P}_\perp (\bm{b}_{\perp};s^{\prime},s)=\bm{0}_\perp,
\end{equation}  
its dipole moment does not vanish in general when the nucleon has a nonzero longitudinal polarization
\begin{equation}
    \int d^{2} b_{\perp}\,
    b_{\perp}^{i}
    \mathcal{P}^{j}_\perp
    (\bm{b}_{\perp};s^{\prime},s)
 = \epsilon^{ij}_\perp\,
    \frac{\sigma_{s^{\prime}s}^{3}}{2}
    \left[J(0)-S(0)\right].
\end{equation}
In particular, the total orbital angular momentum (OAM) about the $z$-axis is given by
\begin{align}
   L_z&:= \int d^{2} b_{\perp}\,
    \left[
    \bm{b}_{\perp}
    \times
    \bm{\mathcal{P}}_{\perp}(\bm{b}_{\perp};s^{\prime},s)
    \right]_{z}\notag\\
&= \sigma^3_{s^{\prime}s}\left[J(0)-S(0)\right].
\end{align}
For a detailed discussion of the longitudinal AM structure, we refer to Ref.~\cite{Lorce:2017wkb}.

In contrast, the 2D spatial distribution of longitudinal flux of transverse momentum (or LT stress) does depend on the nucleon momentum
\begin{equation}
    \Pi^{zi}_\perp (\bm{b}_{\perp},P_z;s^{\prime},s) 
  = - \sigma_{s^{\prime}s}^{3}\,\epsilon^{ij} _\perp
    X_{1}^{j}(\phi_{\bm{b}})
    \Pi^{LT}_{1}
    (b,P_z).
\end{equation}
In this case, the (dipolar) radial distribution is given by
\begin{align}
&   \Pi^{LT}_{1}
    (b,P_z)\notag\\
& = - \frac{1}{2} \frac{d}{db}
    \int \frac{d^{2}\Delta_{\perp}}{\left(2\pi\right)^{2}} \,
    e^{-i\bm{\Delta}_{\perp}\cdot\bm{b}_{\perp}} 
    \,\tilde{\Pi}^{LT}_{1}
    (Q^{2},P_z),
\end{align}
with $\tilde{\Pi}^{LT}_{1}
    (Q^{2},P_z)
=\beta\,\tilde{\mathcal P}_1(Q^2)$. In the non-relativistic limit, the boost parameter $\beta$ does not depend on the momentum transfer, so we have $\Pi^{zi}_\perp (\bm{b}_{\perp},P_z;s^{\prime},s)=\beta\,\mathcal P^i_\perp (\bm{b}_{\perp};s^{\prime},s)$. In the general case, we can write $\beta=\beta_P\,\gamma_P/\gamma$ and the LT stress distribution gets distorted relative to the transverse momentum distribution due to the $\Delta$-dependence of the relativistic factor $\gamma_P/\gamma$. Note, however, that this distortion disappears in the IMF $P_z\to \infty$, so that
\begin{equation}
    \Pi^{zi}_\perp (\bm{b}_{\perp},\infty;s^{\prime},s)= \mathcal P^i_\perp (\bm{b}_{\perp};s^{\prime},s).
\end{equation}

A similar discussion can be made for the transverse energy flux $\mathcal I^i_\perp$ and the transverse flux of longitudinal momentum (or TL stress) $\Pi^{iz}_\perp$, where it suffices to change the sign in front of the intrinsic spin FF $S\mapsto -S$ in the above expressions. For example, the total orbital energy flux about the $z$-axis is given by
\begin{align}
   \bar L_z&:= \int d^{2} b_{\perp}\,
    \left[
    \bm{b}_{\perp}
    \times
    \bm{\mathcal{I}}_{\perp}\left(\bm{b}_{\perp};s^{\prime},s\right)
    \right]_{z}\notag\\
&= \sigma^3_{s^{\prime}s}\left[J(0)+S(0)\right].
\end{align}
The total AM and intrinsic spin about the $z$-axis can then be expressed as $J_z=(L_z+\bar L_z)/2=\sigma^3_{s^{\prime}s}\,J(0)$ and $S_z=(L_z-\bar L_z)/2=\sigma^3_{s^{\prime}s}\,S(0)$, respectively. Similarly, the combination $(\Pi^{zi}_\perp-\Pi^{iz}_\perp)/2$ may be interpreted as an internal torque density, akin to the torsion stress recently discussed in Ref.~\cite{Cosyn:2026gyy}. The latter is intrinsic and requires a state with spin 1 or greater, whereas the one we found here is induced by the motion of the system and exists already for spin-$\frac{1}{2}$ states.

\begin{figure*}[htbp]
  \centering
  \textbf{EF multipole distributions of 2D vector components of the nucleon}

  \vspace{0.2cm} 

  \begin{minipage}{0.37\textwidth}
    \centering
    Quark
  \end{minipage}
  \hspace{-0.1\textwidth}
  \begin{minipage}{0.37\textwidth}
    \centering
    Gluon
  \end{minipage}
  \hspace{-0.1\textwidth}
  \begin{minipage}{0.37\textwidth}
    \centering
    Total
  \end{minipage}

  \begin{minipage}{0.30\textwidth}
    \centering
    \includegraphics[width=\textwidth]{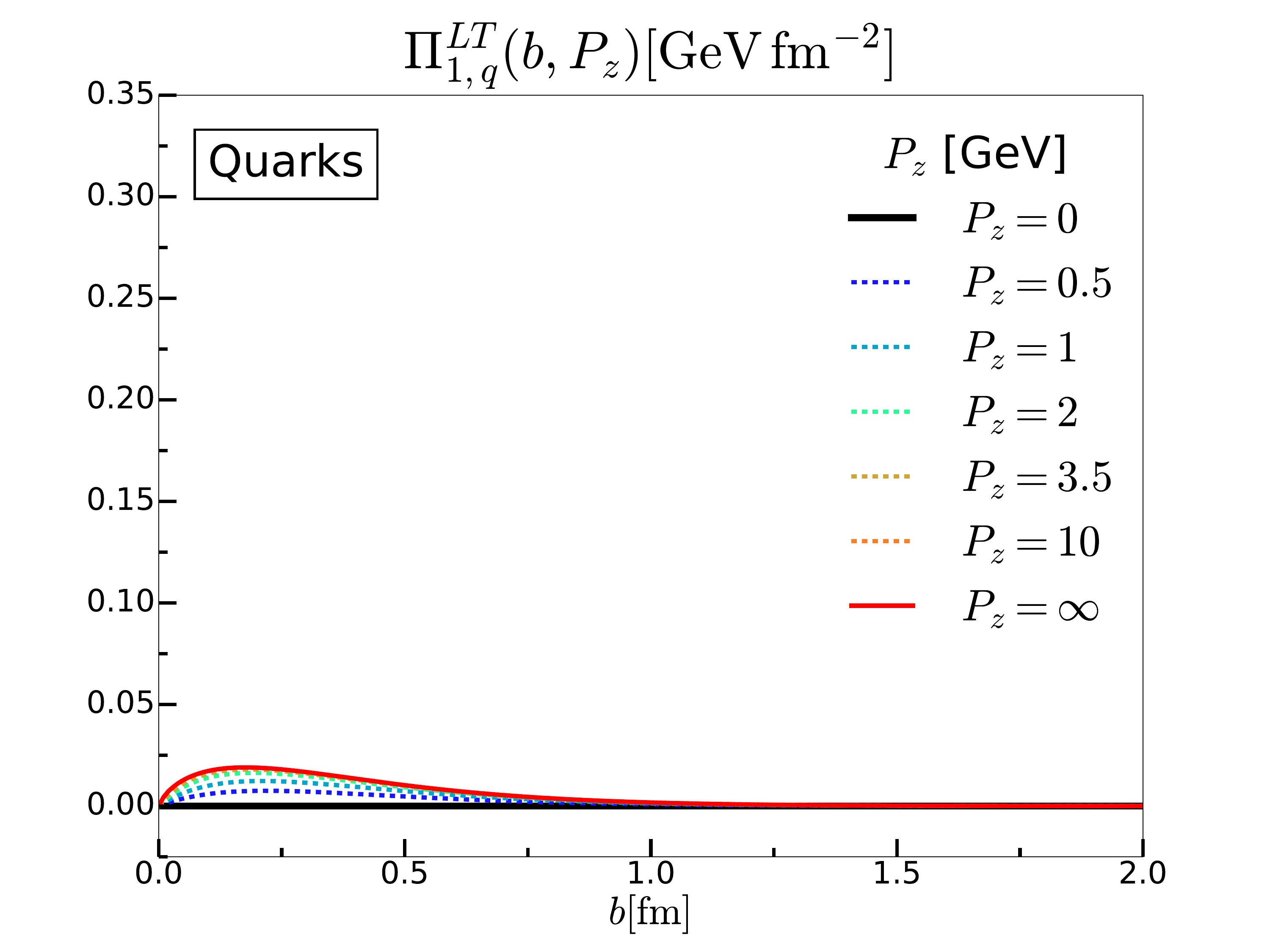}
  \end{minipage}
  \hspace{-0.6cm}
  \begin{minipage}{0.30\textwidth}
    \centering
    \includegraphics[width=\textwidth]{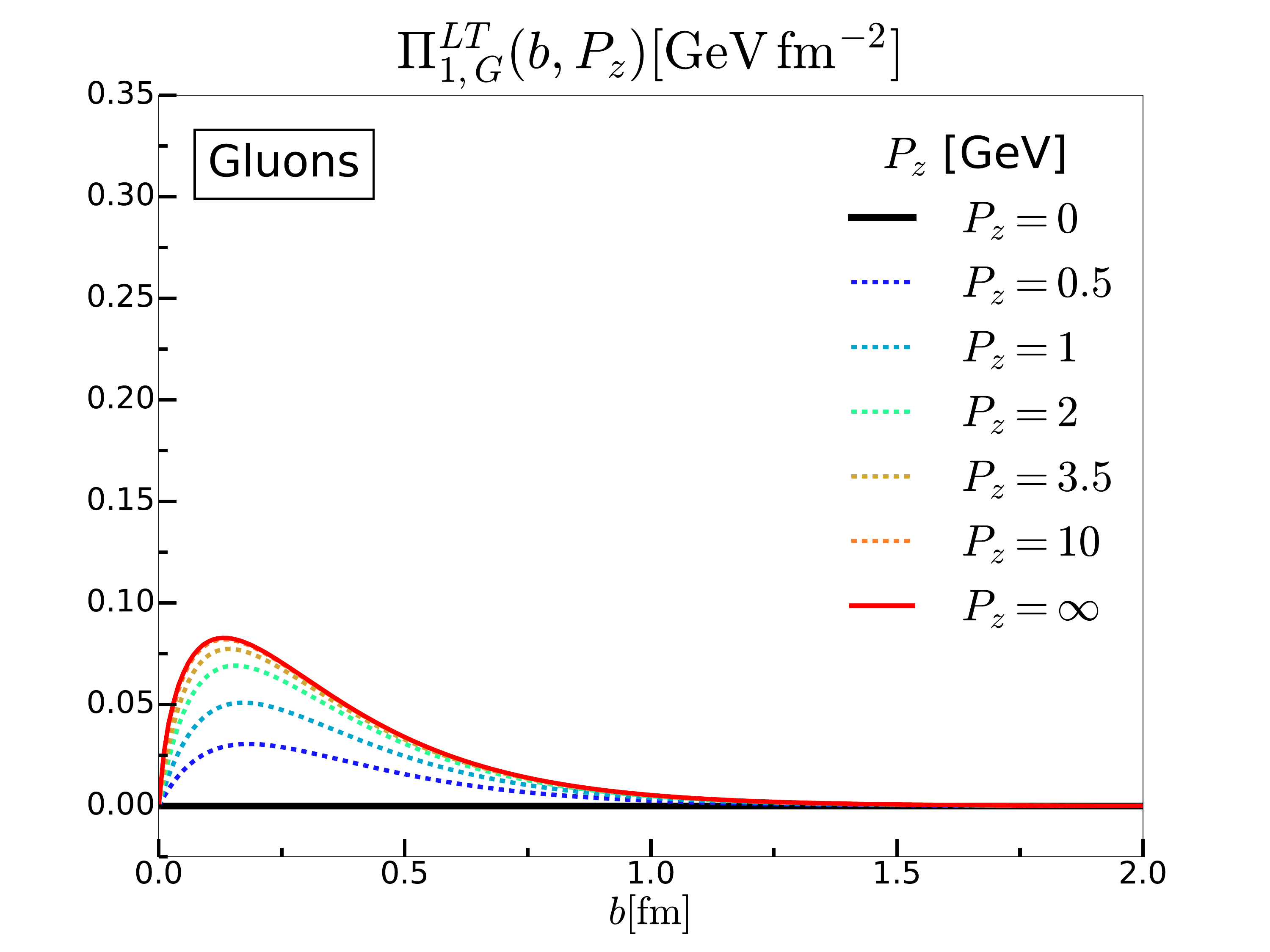}
  \end{minipage}
  \hspace{-0.6cm}
  \begin{minipage}{0.30\textwidth}
    \centering
    \includegraphics[width=\textwidth]{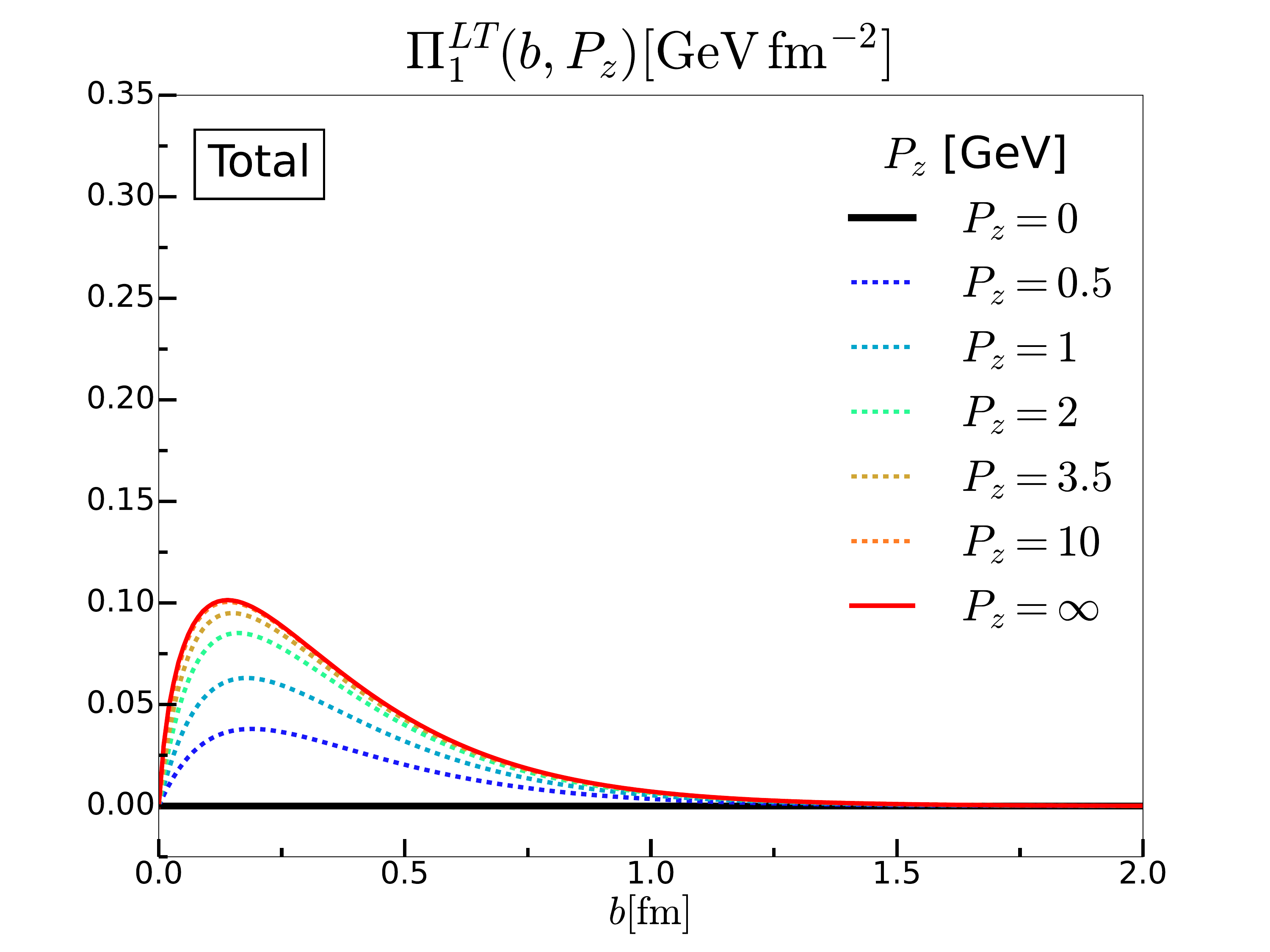}
  \end{minipage}
  
  \vspace{-0cm} 
  
  \begin{minipage}{0.30\textwidth}
    \centering
    \includegraphics[width=\textwidth]{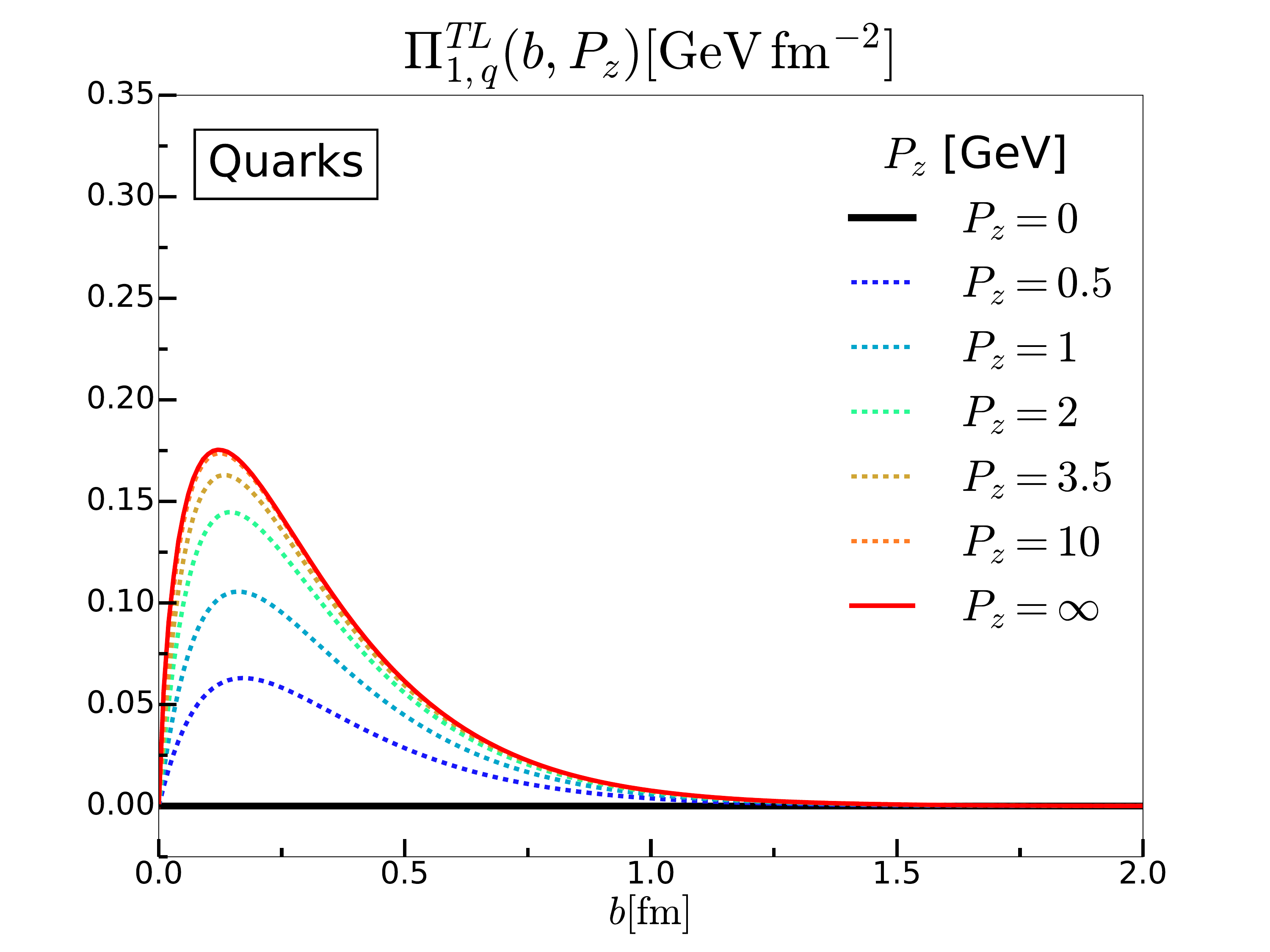}
  \end{minipage}
  \hspace{-0.6cm}
  \begin{minipage}{0.30\textwidth}
    \centering
    \includegraphics[width=\textwidth]{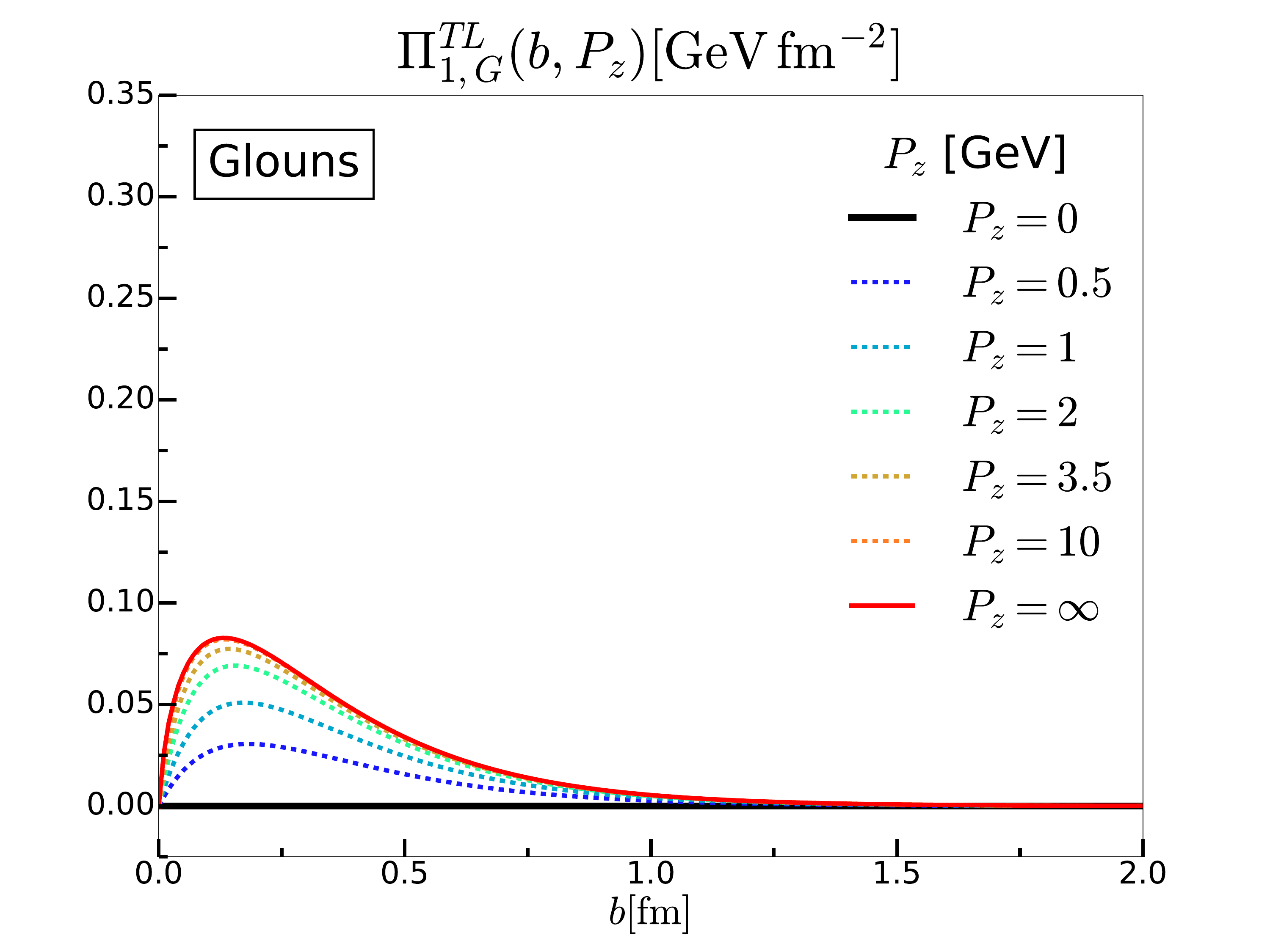}
  \end{minipage}
  \hspace{-0.6cm}
  \begin{minipage}{0.30\textwidth}
    \centering
    \includegraphics[width=\textwidth]{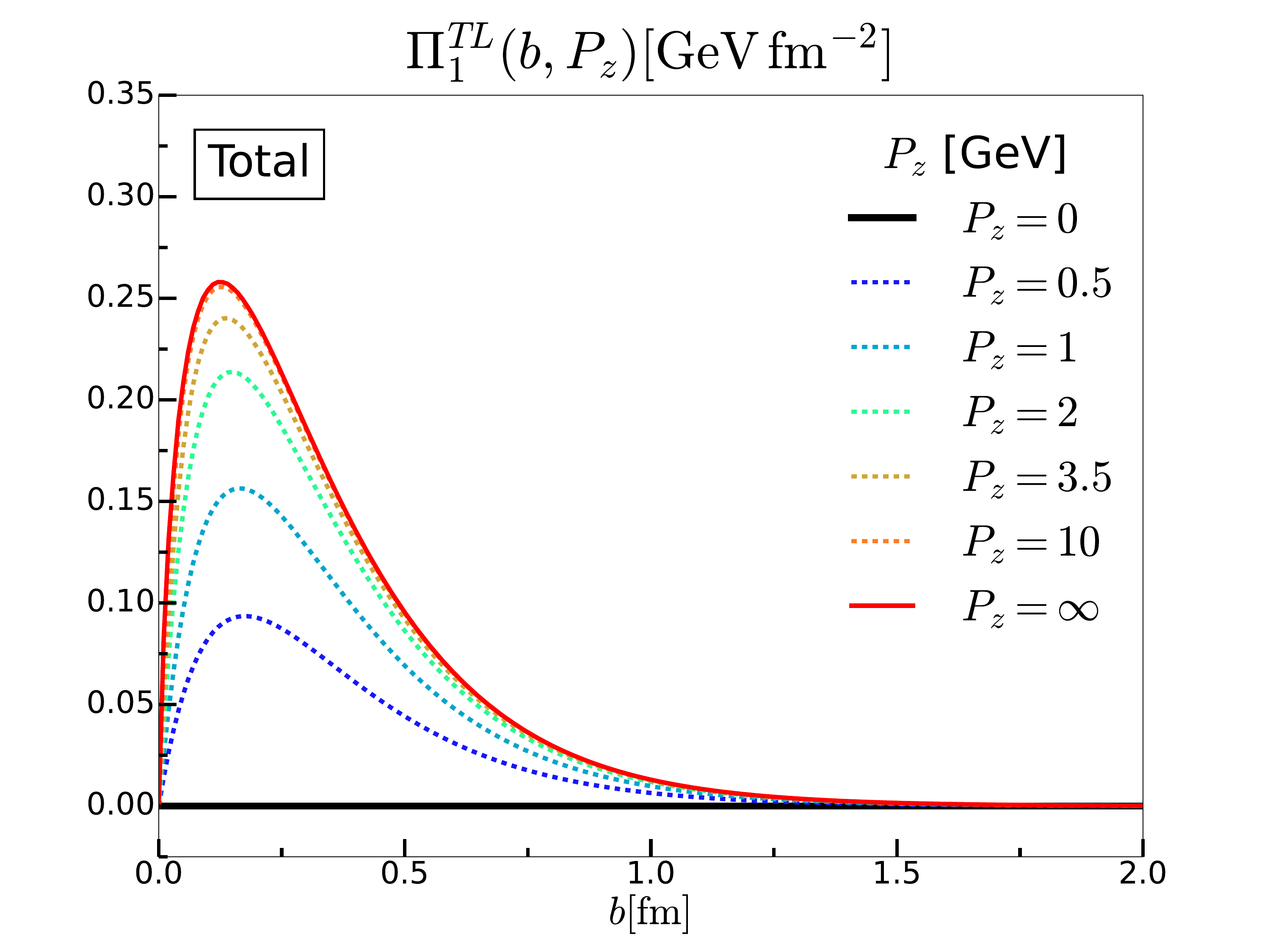}
  \end{minipage}
  
\caption{EF radial distributions in the transverse plane of LT ($\Pi^{zi}_\perp$, upper row) and TL ($\Pi^{iz}_\perp$, lower row) stresses for different values of the nucleon momentum. The first, second, and third columns show the quark, gluon, and total (i.e., quarks + gluons) contributions, respectively. In the IMF (solid red curves), these coincide with the radial distributions of transverse momentum ($\mathcal{P}^{i}_\perp$) and transverse energy flux ($\mathcal{I}^{i}_\perp$), respectively. Based on the simple multipole model of Refs.~\cite{Lorce:2018egm,Won:2025dgc} for the EMT FFs. }

  \label{fig:1}
\end{figure*}
In Fig.~\ref{fig:1}, we show the EF radial distributions of LT and TL stresses for various values of the nucleon momentum $P_z$. Since the phenomenological values of the EMT FFs are not yet well established and our goal is only to illustrate our analytical results, we used the simple multipole model of Refs.~\cite{Lorce:2018egm} and~\cite{Won:2025dgc}. The difference between LT and TL stresses comes from the change of sign for the intrinsic spin FF contribution. For $P_z=0$, i.e.~in the transverse BF, these stresses vanish identically. They are nontrivial when $P_z\neq 0$, and increase in magnitude with $P_z$. They reach their maximal value for $P_z\to\infty$, i.e.~in the IMF, where they become equal to the dipolar radial distributions of transverse momentum $\mathcal{P}_1(b)$ and transverse energy flux $\mathcal{I}_1(b)$, respectively.

\begin{figure*}[htbp]
  \centering
  \textbf{Total EF distributions of 2D vector distributions for a longitudinally polarized nucleon}
  \vspace{-0.2cm} 
  \includegraphics[width=0.90\textwidth]{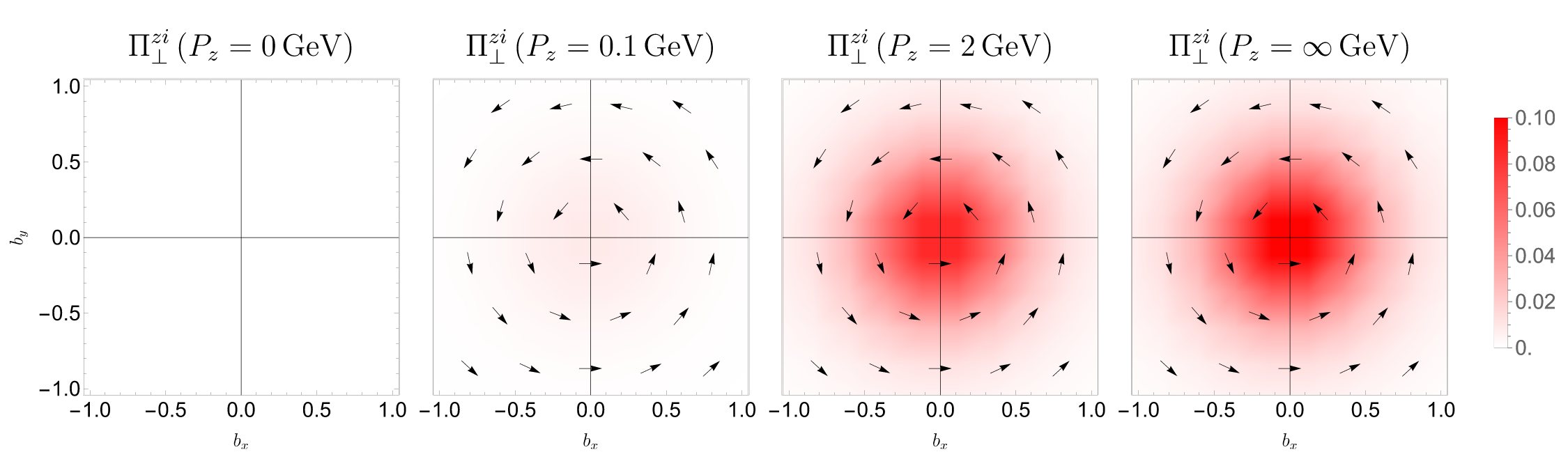}
  \vspace{-0.2cm}
  \includegraphics[width=0.90\textwidth]{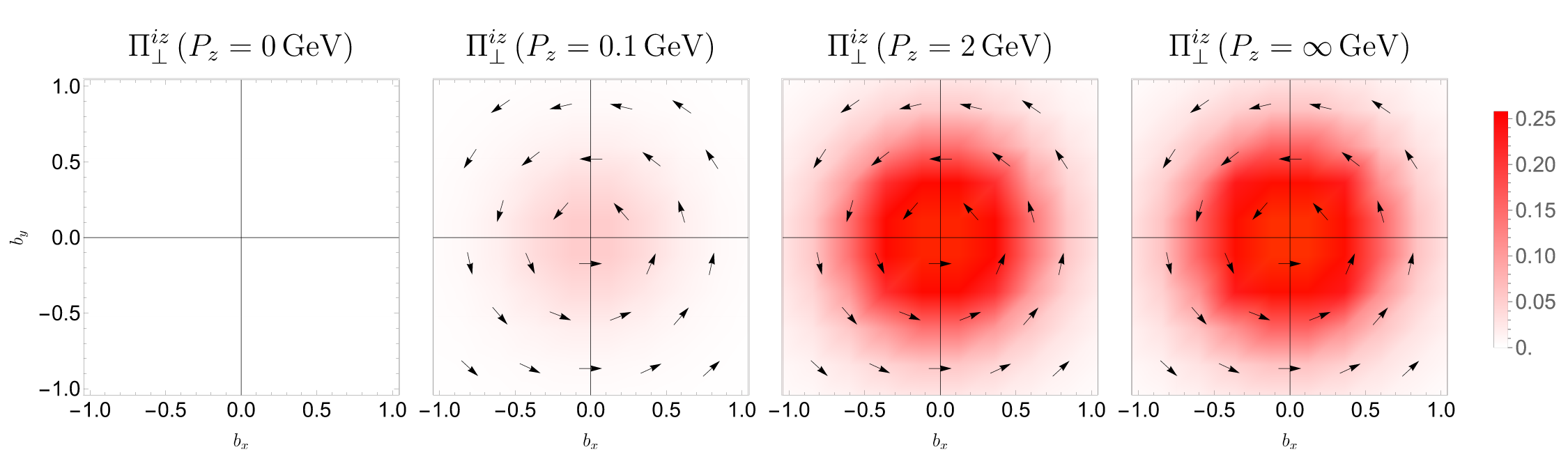}
  \caption{EF distributions of total (i.e., quarks + gluons) LT ($\Pi_{\perp}^{zi}$, first row) and TL ($\Pi_{\perp}^{iz}$, second row) stresses for a longitudinally polarized nucleon and different values of the nucleon momentum. 
  The direction of $\Pi^{zi}_\perp$ and  $\Pi^{iz}_\perp$ at a point $\bm b_\perp$ in the transverse plane is represented by an arrow, and the magnitude by a color scale. Based on the simple multipole model of Refs.~\cite{Lorce:2018egm,Won:2025dgc} for the EMT FFs. 
  }

  \label{fig:2}
\end{figure*} 
In Fig.~\ref{fig:2}, we represent the total EF distributions of LT and TL stresses in a longitudinally polarized nucleon as vector fields in the transverse plane. 
The magnitudes of these vector fields are encoded by the overlayed density plots. We clearly see the characteristic orbital structure $\epsilon^{ij} _\perp X_{1}^{j}(\phi_{\bm{b}})$ of these distributions.

\subsection{Tensor distributions $\Pi^{ij}_\perp$}

We now turn our attention to the tensor distributions. The transverse stress tensor can be decomposed into pure trace and traceless contributions
\begin{align}
    \Pi_{\perp}^{ij}
  = \delta_{\perp}^{ij}\,
    \sigma 
  + \Sigma^{ij}_\perp,
\label{force_field_distribution_2}
\end{align}
where $\sigma$ and $\Sigma^{ij}_\perp$ denote the 2D isotropic and anisotropic stress tensor distributions, respectively. In other words, we can write (with implicit sum over repeated indices)
\begin{subequations}
\begin{align}
    \sigma 
&= \frac{1}{2}
    \,\Pi_{\perp}^{kk}, 
\\
    \Sigma^{ij}_\perp&=
 \Pi^{<ij>}_\perp:= \Pi^{ij}_\perp
  - \frac{1}{2} 
    \,\delta_{\perp}^{ij}
    \,\Pi^{kk}_\perp.
\label{def_stress_tensor_distribution}
\end{align}
\end{subequations}
Equivalently, the component $\Pi^{ij}_\perp$ can be interpreted as the flux in the $i$-direction of the $j$th component of momentum. The absence of an antisymmetric contribution in Eq.~\eqref{force_field_distribution_2} is a consequence of the fact that $\Delta_z=0$ in EF.

\subsubsection{Isotropic stress $\sigma$}

In contrast to vector components, the tensor EMT components involve only transverse indices, and therefore do not mix under a longitudinal boost. However, they undergo a Wigner spin rotation, in addition to the distortion arising from the relativistic normalization factor in $\langle\langle
    T^{ij}
    \rangle\rangle$.
We then find that the matrix elements of the (transverse) isotropic stress are given in the EF by
\begin{align}
&   \hspace{-0.5cm}
    \frac{1}{2} 
    \langle\langle
    T^{kk}
    \rangle\rangle_{\mathrm{EF}}  
    \notag\\
&   \hspace{-0.5cm}
  = \frac{M}{\gamma}\left[
    \delta_{s^{\prime}s}
    \cos{\theta}
  + i
    \epsilon^{kl}_\perp
    \sigma_{s^{\prime}s}^{k}\,
        X_{1}^{l}(\phi_{\bm{\Delta}})\,
    \sin{\theta}
    \right] F_{0}(Q^{2}).
\end{align}
The corresponding 2D spatial distribution can be decomposed as follows
\begin{align}
&    \sigma
    (\bm{b}_{\perp},P_{z};s^{\prime},s) \notag\\
& = \delta_{s^{\prime}s}\,
    \sigma_{0}
    (b,P_{z})
 + \epsilon^{kl}_\perp
    \sigma_{s^{\prime}s}^{k}\,
    X_{1}^{l}(\phi_{\bm{b}})\,
    \sigma_{1}
    (b,P_{z}),
    \label{isotropic_stress}
\end{align}
where the monopolar and dipolar radial distributions are given by
\begin{align}
    \hspace{-0.7cm}
    \sigma_{0} 
    (b,P_{z})
& = M 
    \int \frac{d^{2}\Delta_{\perp}}{\left(2\pi\right)^{2}} \,
    e^{-i\bm{\Delta}_{\perp}\cdot\bm{b}_{\perp}} 
    \,\tilde{\sigma}_{0}
    (Q^{2},P_{z}),\\ 
    \hspace{-0.7cm}
    \sigma_{1} 
    (b,P_{z})
& = - \frac{1}{2}
    \frac{d}{db}
    \int \frac{d^{2}\Delta_{\perp}}{\left(2\pi\right)^{2}} \,
    e^{-i\bm{\Delta}_{\perp}\cdot\bm{b}_{\perp}} 
    \,\tilde{\sigma}_{1}
    (Q^{2},P_{z})
\end{align}
with
\begin{align}
    \tilde{\sigma}_{0}
    (Q^{2},P_{z})
& = \frac{\gamma_{P}}{\gamma}
    \cos{\theta}
   \, F_{0}
    (Q^{2}),    \\
    \tilde{\sigma}_{1}
    (Q^{2},P_{z})
& = \frac{\gamma_{P}}{\gamma}
    \frac{\sin{\theta}}{\sqrt{\tau}}
    \,F_{0}
    (Q^{2}).
\end{align} 

While the isotropic stress distribution depends on $P_z$, its integral over the transverse plane does not
\begin{equation}\label{integrated_isotropic}
    \int d^{2}b_{\perp}\,
    \sigma
    (\bm{b}_{\perp},P_{z};s^{\prime},s) 
 = \delta_{s^{\prime}s}\,
    F_{0}(0).
\end{equation}
Summing over all the partons, it follows from Eq.~\eqref{symmetry} that the total isotropic stress vanishes. This result is nothing more than the 2D version of the von Laue condition~\cite{Laue:1911lrk} that expresses global equilibrium~\cite{Polyakov:2018zvc,Lorce:2018egm,Lorce:2021xku}.

\begin{figure*}[htbp]
  \centering
  \textbf{EF multipole distributions of 2D isotropic stress of the nucleon}

  \vspace{0.2cm} 

  \begin{minipage}{0.37\textwidth}
    \centering
    Quark
  \end{minipage}
  \hspace{-0.1\textwidth}
  \begin{minipage}{0.37\textwidth}
    \centering
    Gluon
  \end{minipage}
  \hspace{-0.1\textwidth}
  \begin{minipage}{0.37\textwidth}
    \centering
    Total
  \end{minipage}

  \begin{minipage}{0.30\textwidth}
    \centering
    \includegraphics[width=\textwidth]{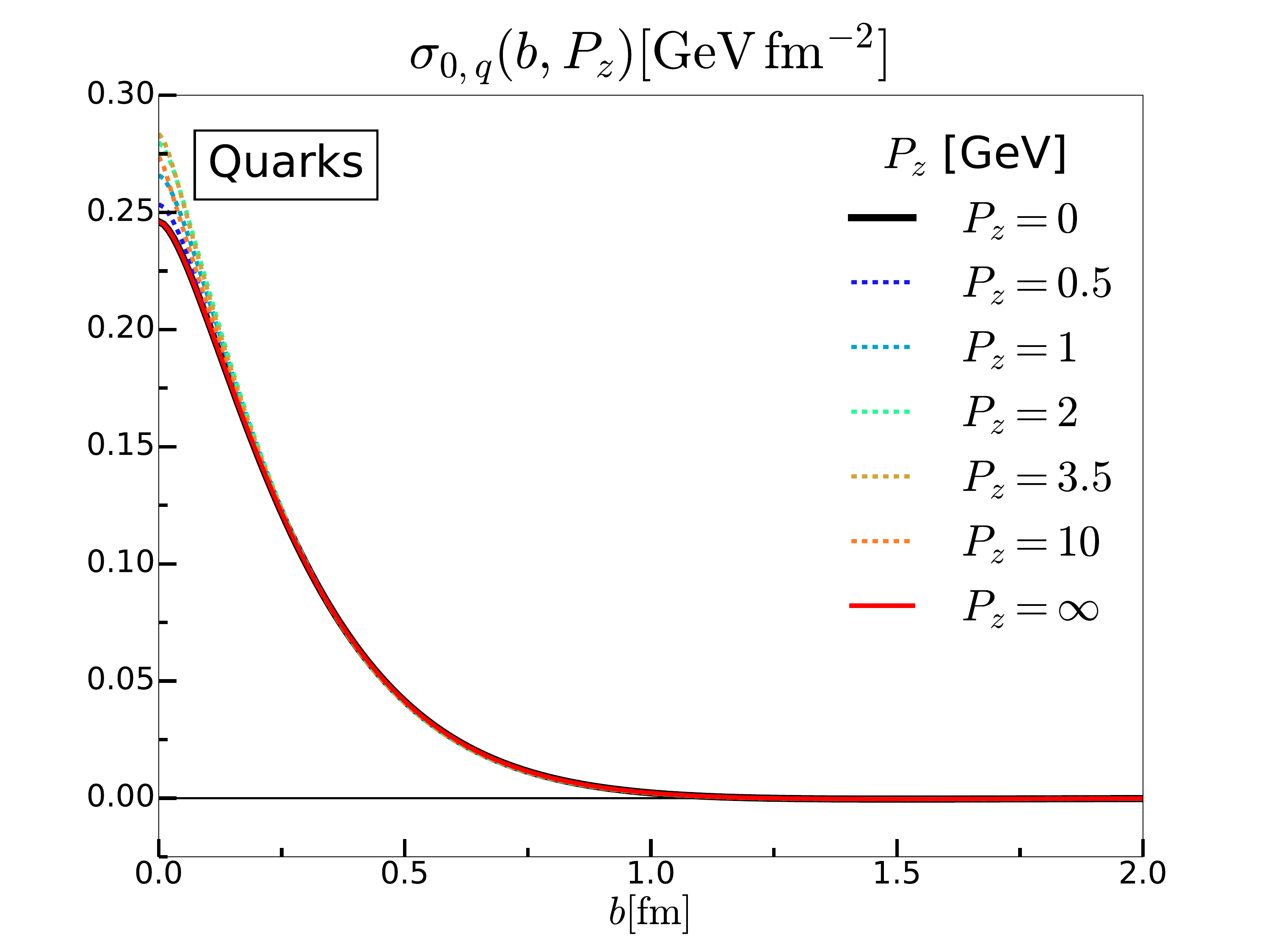}
  \end{minipage}
  \hspace{-0.6cm}
  \begin{minipage}{0.30\textwidth}
    \centering
    \includegraphics[width=\textwidth]{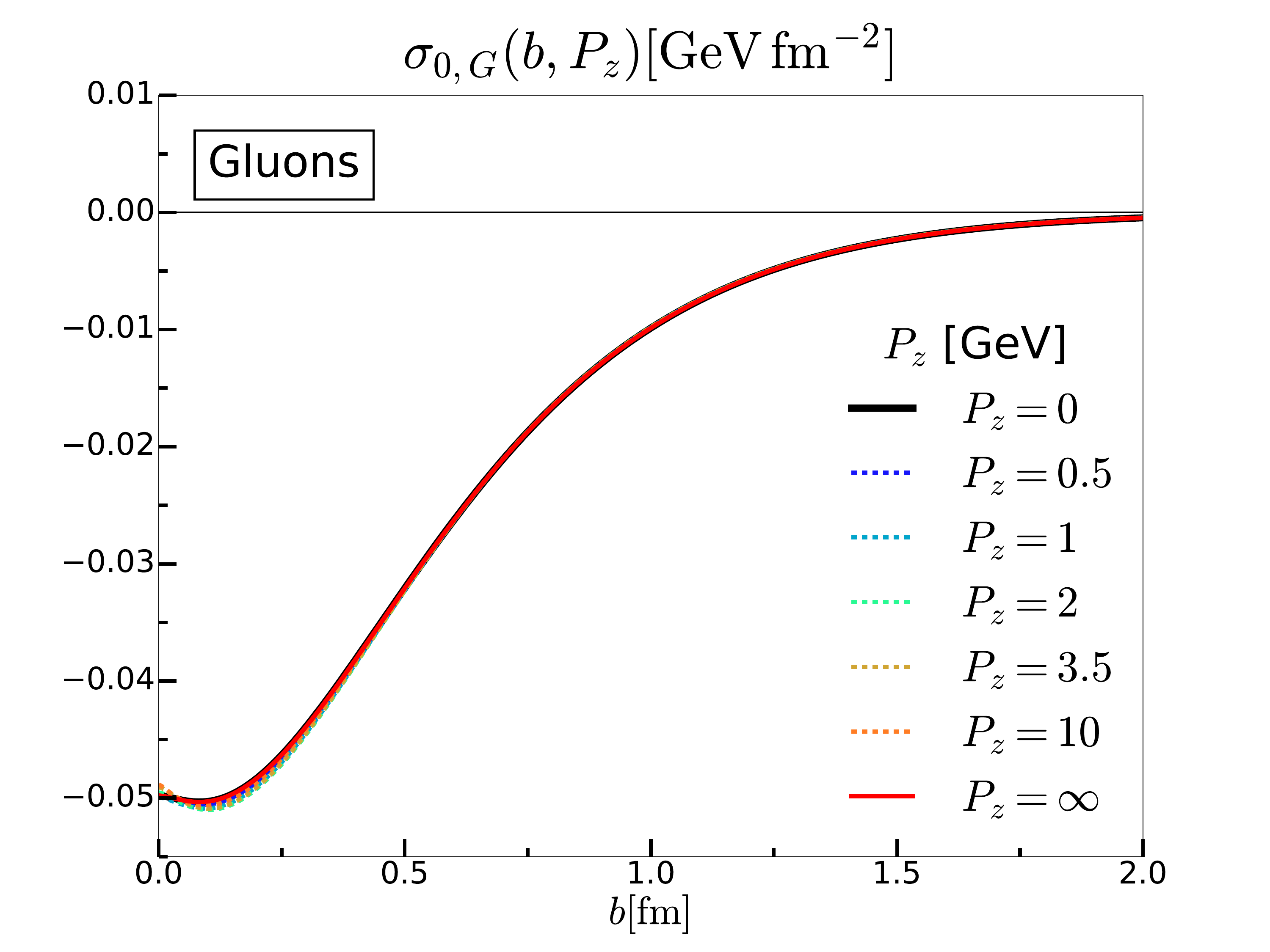}
  \end{minipage}
  \hspace{-0.6cm}
  \begin{minipage}{0.30\textwidth}
    \centering
    \includegraphics[width=\textwidth]{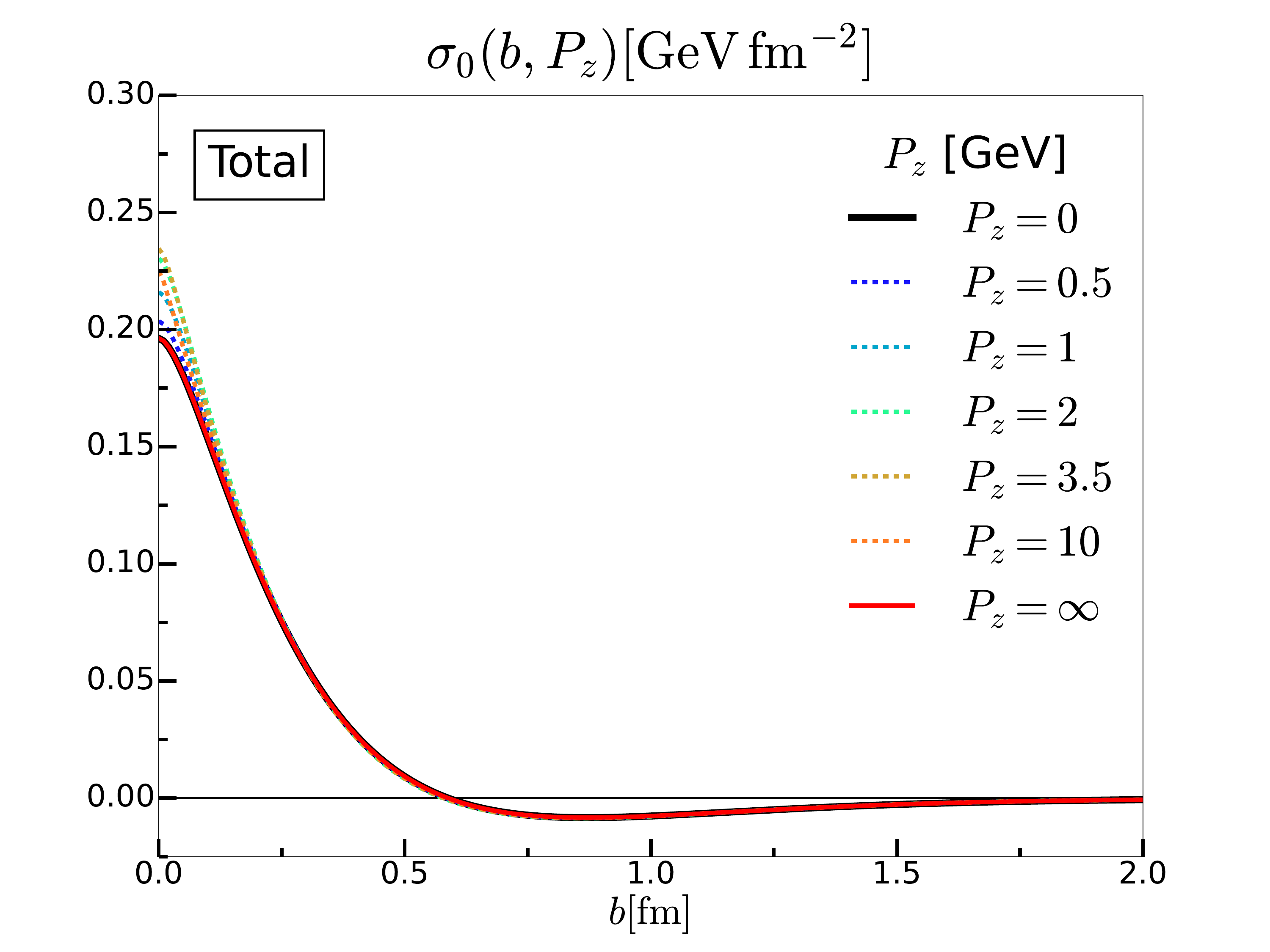}
  \end{minipage}
  
  \vspace{-0.0cm} 
  
  \begin{minipage}{0.30\textwidth}
    \centering
    \includegraphics[width=\textwidth]{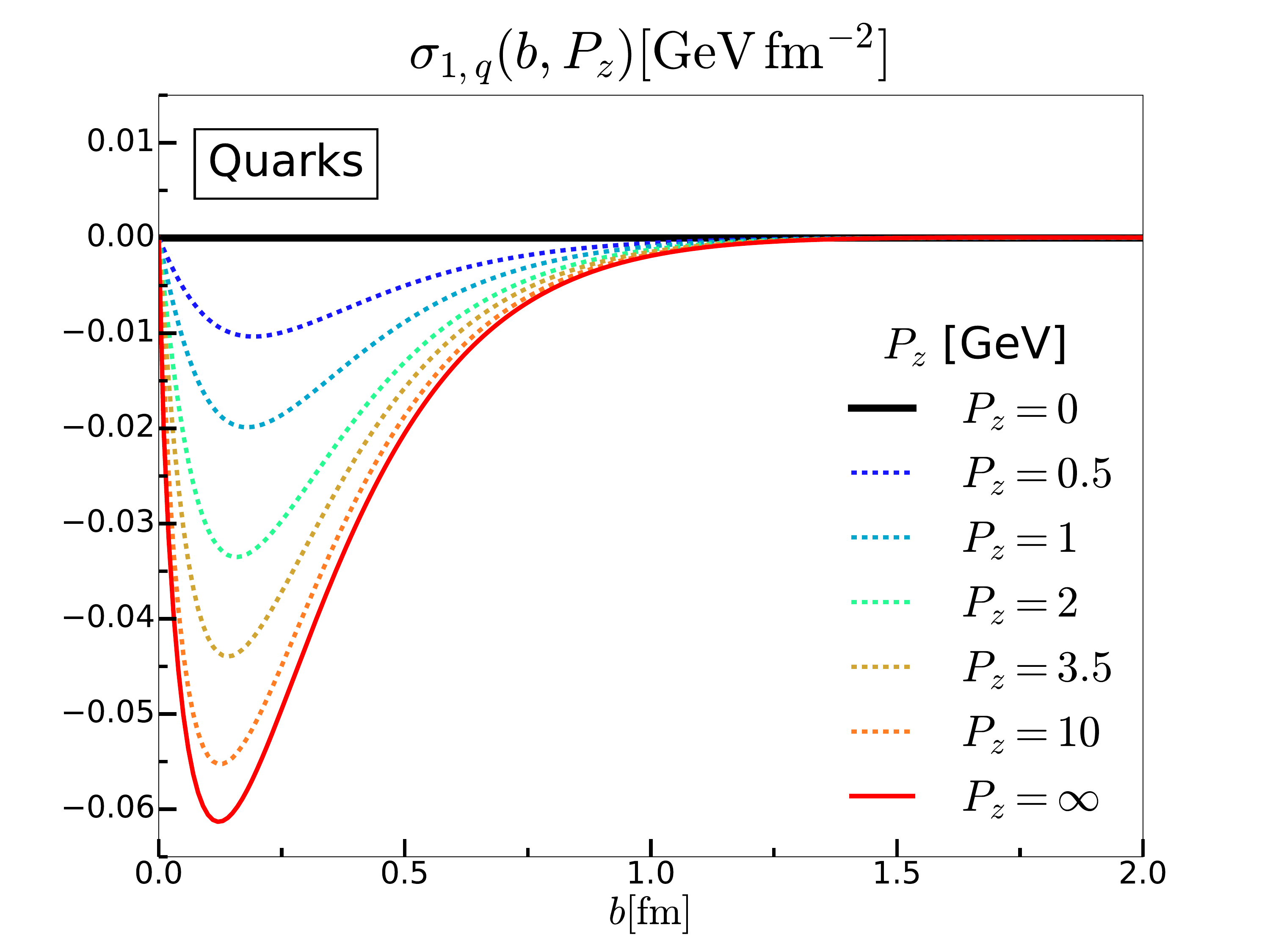}
  \end{minipage}
  \hspace{-0.6cm}
  \begin{minipage}{0.30\textwidth}
    \centering
    \includegraphics[width=\textwidth]{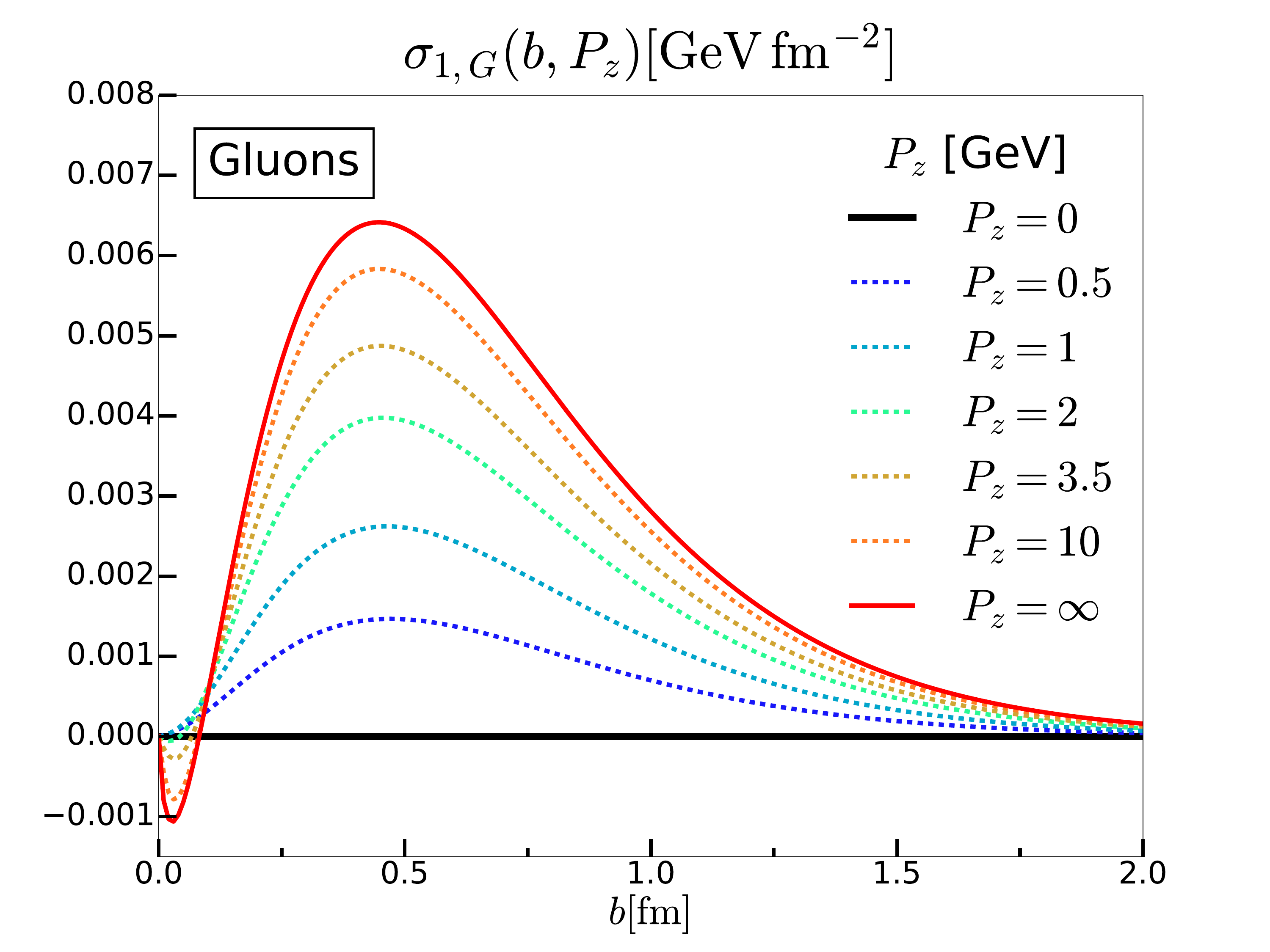}
  \end{minipage}
  \hspace{-0.6cm}
  \begin{minipage}{0.30\textwidth}
    \centering
    \includegraphics[width=\textwidth]{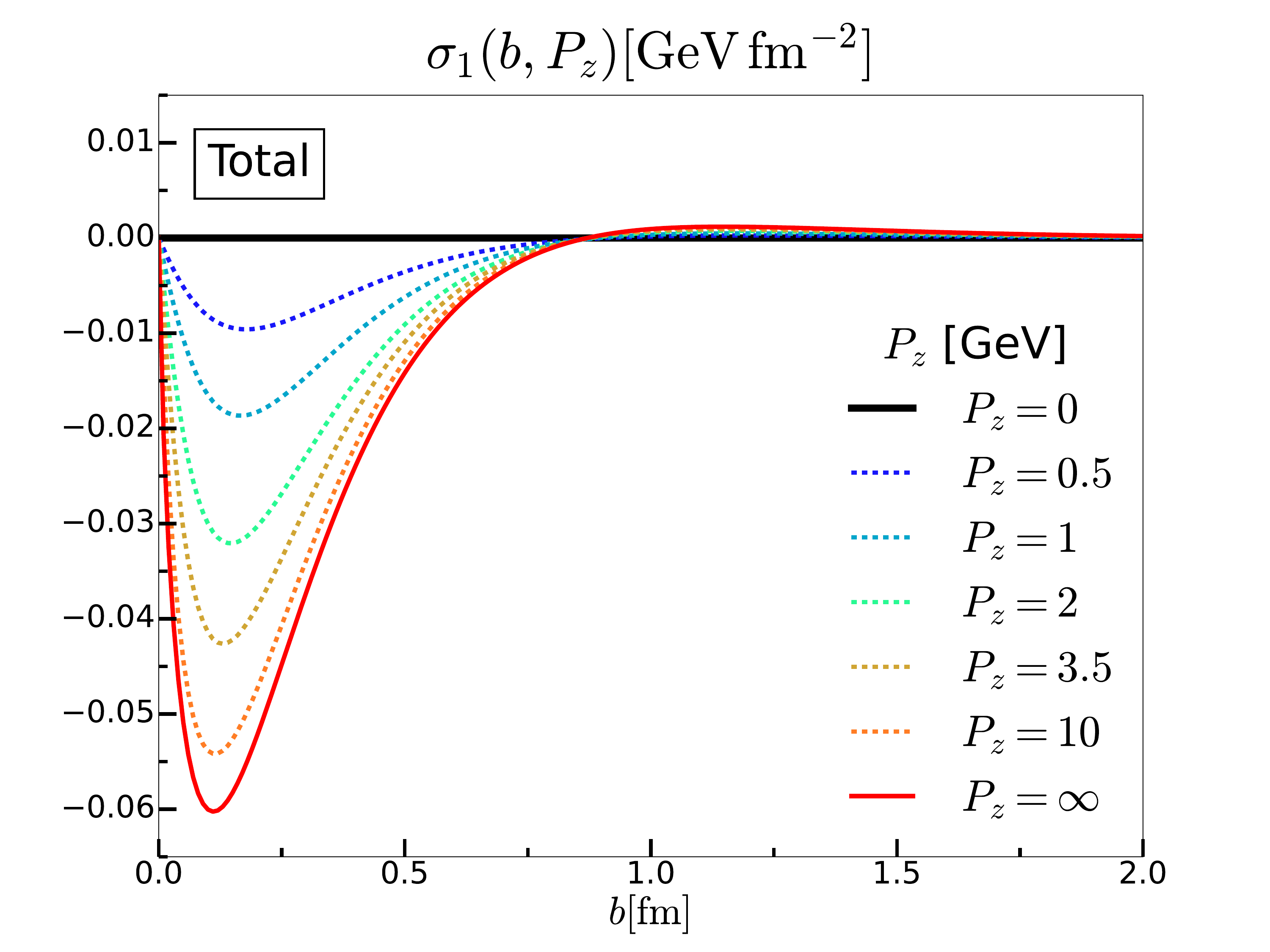}
  \end{minipage}
  
  \caption{EF monopolar (upper row) and dipolar (lower row) radial distributions in the transverse plane of 2D isotropic stress in the nucleon for different values of the nucleon momentum. The first, second, and third columns show the quark, gluon, and total (i.e., quarks + gluons) contributions, respectively. Based on the simple multipole model of Refs.~\cite{Lorce:2018egm,Won:2025dgc} for the EMT FFs.}
  \label{fig:3}
\end{figure*}
In Fig,~\ref{fig:3}, we show the monopolar and dipolar radial distributions of isotropic stress for various values of the nucleon momentum. It appears that the quark contribution dominates over the gluon one throughout the transverse plane. The reason is that, contrary to their quark counterparts, the gluon EMT FFs $D_G$ and $\bar C_G$ have opposite signs in the multipole model of Refs.~\cite{Lorce:2018egm,Won:2025dgc}, and thus largely cancel each other.

\begin{figure*}[htbp]
  \centering
  \textbf{Total EF distributions of 2D isotropic stress for a transversely polarized nucleon}

  \vspace{0.5cm} 

  \includegraphics[width=0.90\textwidth]{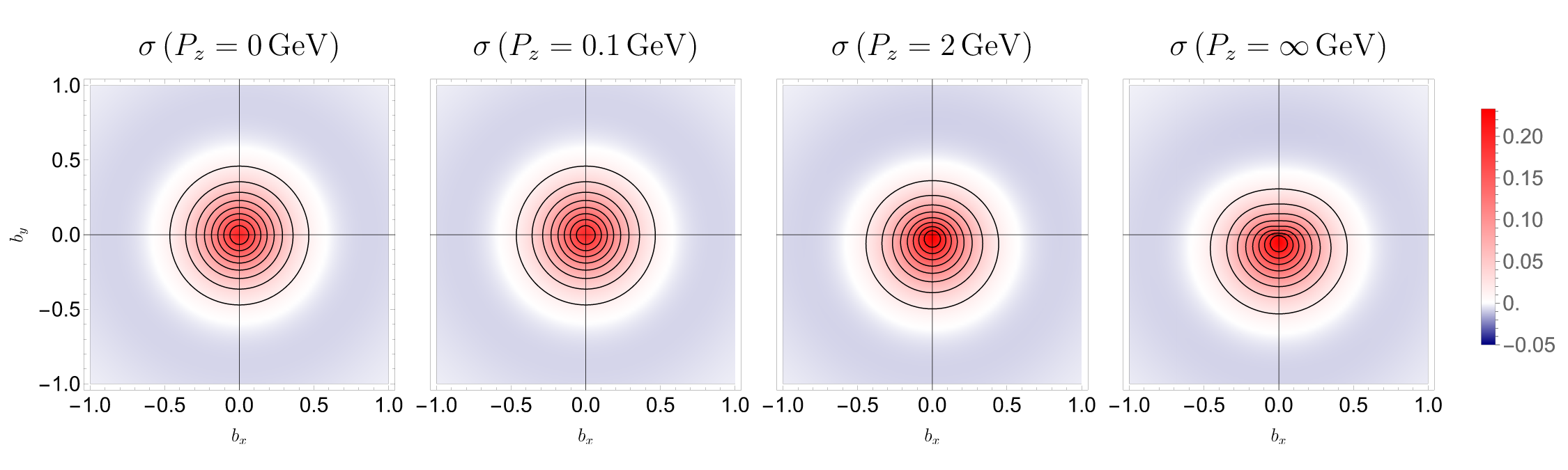}



  \caption{
EF distributions of total (i.e., quarks + gluons) transverse isotropic stress in the transverse plane for a nucleon polarized along the $x$-axis and different values of the nucleon momentum. Based on the simple multipole model of Refs.~\cite{Lorce:2018egm,Won:2025dgc} for the EMT FFs.
  }

  \label{fig:4}
\end{figure*} 
In Fig.~\ref{fig:4}, we represent the total EF distribution of transverse isotropic stress in a transversely polarized nucleon. We choose the transverse polarization along the $x$-axis, i.e.~$\ket{s_{x}=\pm1/2} = \left( \ket{s = 1/2} \pm \ket{s = -1/2} \right) / \sqrt{2}$. The dipolar distortion along the $y$-axis, which increases with larger values of $P_z$, is the result of the Wigner spin rotation. This shift goes in the opposite direction compared to the one observed in the proton electric charge distribution~\cite{Chen:2022smg,Kim:2021kum} and the scalar EMT distributions~\cite{Won:2025dgc}. This can be traced back to the fact that the EMT multipole FF $F_0$ is always negative in the model we used to illustrate our results.

\subsubsection{Anisotropic stress tensor $\Sigma^{ij}_\perp$}

Finally, we find that the EF matrix elements of the (transverse) anisotropic stress tensor are given by
\begin{widetext}
\begin{align}
    \langle\langle
    T^{\langle ij\rangle}
    \rangle\rangle_{\mathrm{EF}}
   &= \frac{M}{\gamma}\left[
    \delta_{s^{\prime}s}
    \cos{\theta}
  + i
    \epsilon^{kl}_\perp
    \sigma_{s^{\prime}s}^{k}\,
        X_{1}^{l}(\phi_{\bm{\Delta}})
    \sin{\theta}
    \right] \tau X_{2}^{ij}(\phi_{\bm{\Delta}})\,F_{2}(Q^{2})\notag\\
& = \frac{M}{\gamma}
    \left[
     \delta_{s^{\prime}s}\,
    X_{2}^{ij}(\phi_{\bm{\Delta}})
    \cos{\theta}+i 
    \epsilon^{kl}_\perp
    \sigma_{s^{\prime}s}^{k}\left(
    \frac{1}{2} P^{ij,lm}_\perp
    X_{1}^{m}(\phi_{\bm{\Delta}})
   +X_{3}^{ijl}(\phi_{\bm{\Delta}})
  \right)\sin{\theta}\right]\tau F_2(Q^2), 
    \label{Tij} 
\end{align}
\end{widetext}
where $P^{ij,lm}_\perp=\left(\delta^{il}_\perp\delta^{jm}_\perp+\delta^{im}_\perp\delta^{jl}_\perp-\delta^{ij}_\perp\delta^{lm}_\perp\right)/2$ is the rank-2 symmetric traceless projector. In the first line, we clearly see the effect of the Wigner rotation. In the second line, we decomposed the EF matrix elements into 2D multipole contributions. The corresponding 2D anisotropic stress tensor distribution can then be expressed as
\begin{widetext}
\begin{equation}
    \Sigma^{ij}_\perp
    (\bm{b}_{\perp},P_{z};s^{\prime},s)
 = \epsilon^{kl}_\perp  
    \sigma_{s^{\prime}s}^{k}\,\frac{1}{2} P_{\perp}^{ij,lm} 
    X_{1}^{m} (\phi_{\bm{b}})\, 
    \Sigma_{1} (b,P_{z})
 + \delta_{s^{\prime}s}\,
    X_{2}^{ij}(\phi_{\bm{b}})\,
    \Sigma_{2} (b,P_{z})
 + \epsilon^{kl}_\perp  
    \sigma_{s^{\prime}s}^{k}\,
    X_{3}^{lij}(\phi_{\bm{b}}) \,
    \Sigma_{3} (b,P_{z}),
\label{anisotropic_stress_tensor}
\end{equation}
\end{widetext}
where the multipolar radial distributions are defined as 
\begin{align}
&   \Sigma_{n} 
    (b,P_{z})
  = i^{n\,\mathrm{mod}\,2}\,
     M \left(\frac{ib}{2M}\right)^{n} \notag\\
&   \hspace{0.3cm} 
    \times
    \left[
    \frac{1}{b}
    \frac{d}{db}
    \right]^{n}
    \int \frac{d^{2}\Delta_{\perp}}{\left(2\pi\right)^{2}} \,
    e^{-i\bm{\Delta}_{\perp}\cdot\bm{b}_{\perp}} 
    \tilde{\Sigma}_{n}
    (Q^{2},P_{z})
\end{align}
with the amplitudes
\begin{align}
    \hspace{-.5cm}
    \tilde{\Sigma}_{1}
    (Q^{2},P_{z})&=\tau\tilde{\Sigma}_{3}
    (Q^{2},P_{z})
 = \frac{\gamma_{P}}{\gamma}
    \frac{\sin{\theta}}{\sqrt{\tau}} \,
    \tau
    F_{2}
    (Q^{2}),  \notag\\
    \hspace{-.5cm}
    \tilde{\Sigma}_{2}
    (Q^{2},P_{z})
& = \frac{\gamma_{P}}{\gamma}
    \cos{\theta}\,
    F_{2}
    (Q^{2}).
\end{align} 

Like the monopole moment of isotropic stress~\eqref{integrated_isotropic}, the quadrupole moment of anisotropic stress
\begin{align}
&\int d^{2}b_{\perp}\,
    2\left|\bm{b}_{\perp}\right|^{2}X_{2}^{ij}(\phi_{\bm{b}})\,
    \Sigma^{ij}_{\perp}(\bm{b}_{\perp},P_{z};s',s) \notag\\
&   =\delta_{s^{\prime}s}\int d^{2}b_{\perp}\,
    \left|\bm{b}_{\perp}\right|^{2}
    \Sigma_{2}
    (b,P_{z}) \notag\\
 & = - \frac{2}{M}\,
    \delta_{s^{\prime}s}\,
    F_{2}
    (0)
\end{align}
depends neither on $P_z$ nor on the nucleon polarization.

\begin{figure*}[htbp]
  \centering
  \textbf{EF multipole distributions of 2D anisotropic stress of the nucleon}

  \vspace{0.2cm} 

  \begin{minipage}{0.37\textwidth}
    \centering
    Quark
  \end{minipage}
  \hspace{-0.1\textwidth}
  \begin{minipage}{0.37\textwidth}
    \centering
    Gluon
  \end{minipage}
  \hspace{-0.1\textwidth}
  \begin{minipage}{0.37\textwidth}
    \centering
    Total
  \end{minipage}

  \begin{minipage}{0.30\textwidth}
    \centering
    \includegraphics[width=\textwidth]{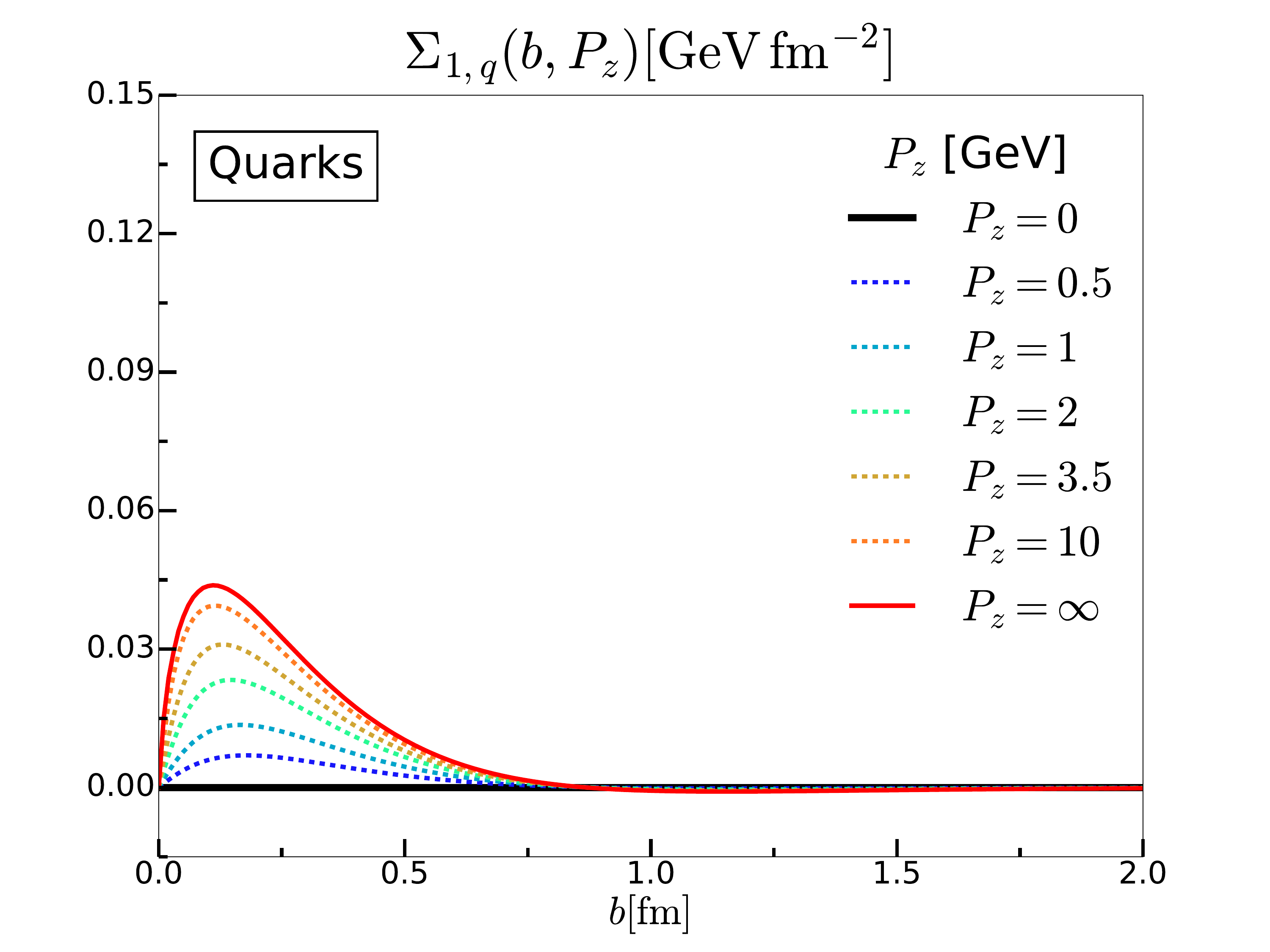}
  \end{minipage}
  \hspace{-0.6cm}
  \begin{minipage}{0.30\textwidth}
    \centering
    \includegraphics[width=\textwidth]{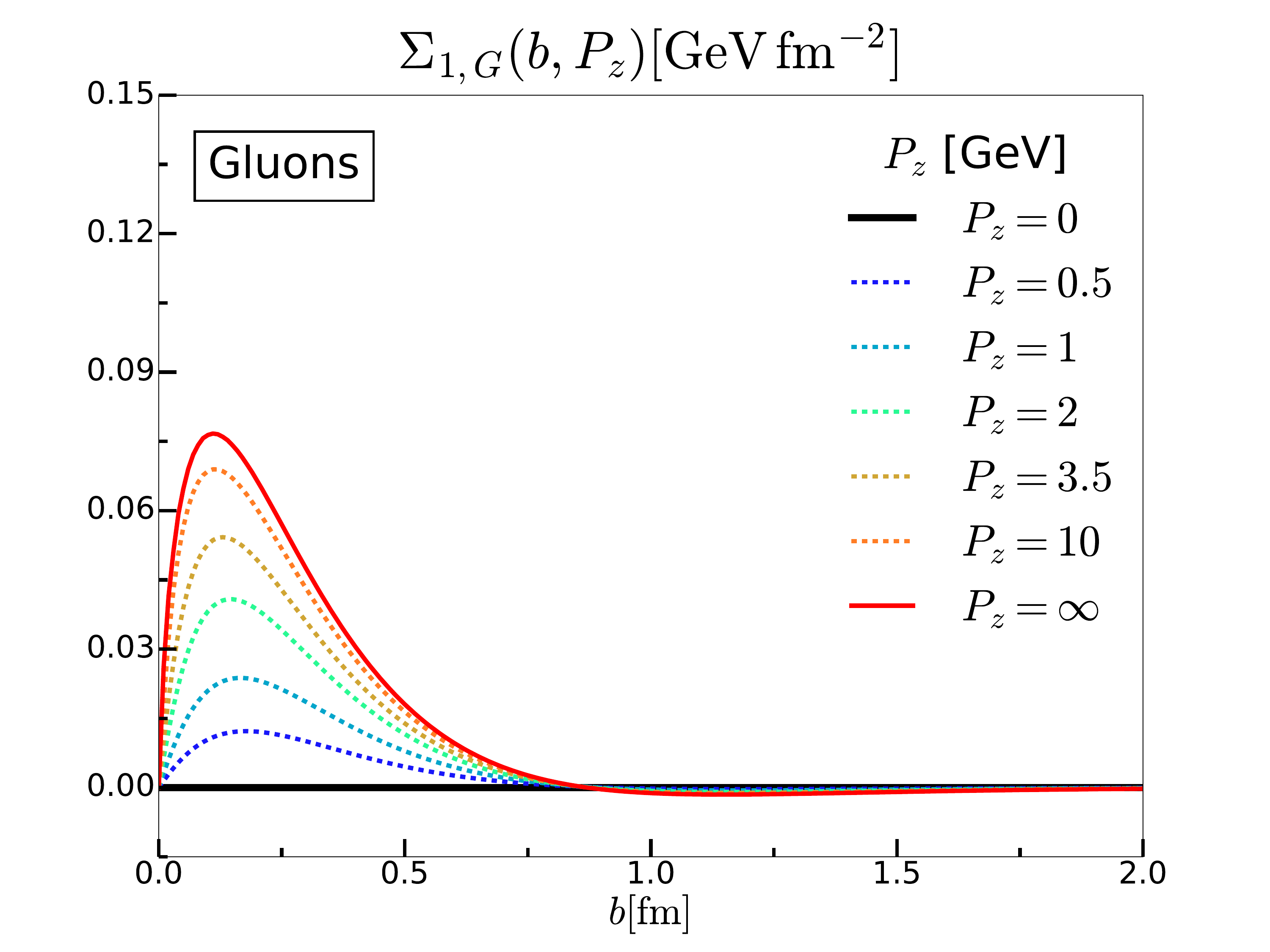}
  \end{minipage}
  \hspace{-0.6cm}
  \begin{minipage}{0.30\textwidth}
    \centering
    \includegraphics[width=\textwidth]{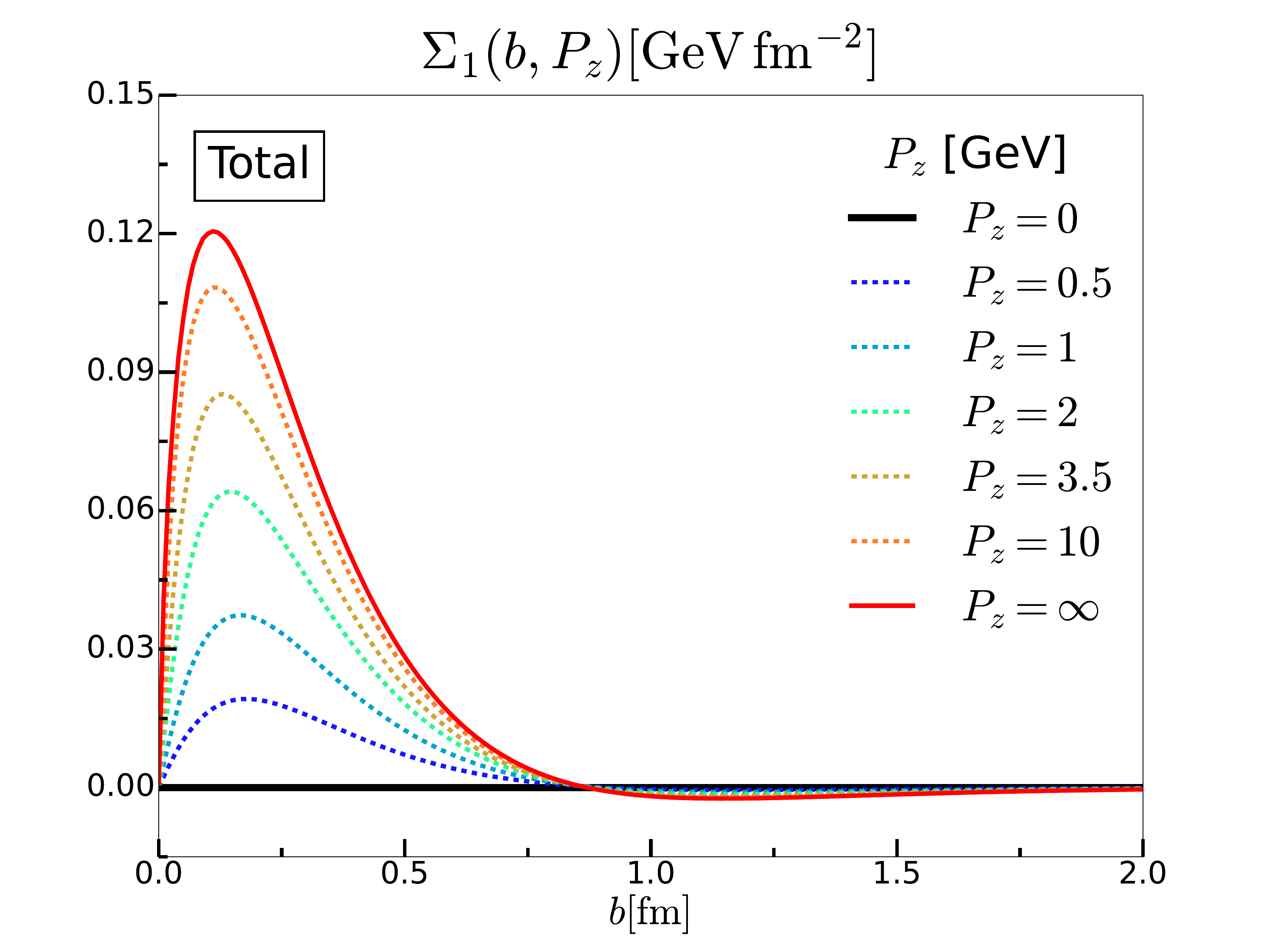}
  \end{minipage}
  
  \vspace{-0.0cm} 
  
  \begin{minipage}{0.30\textwidth}
    \centering
    \includegraphics[width=\textwidth]{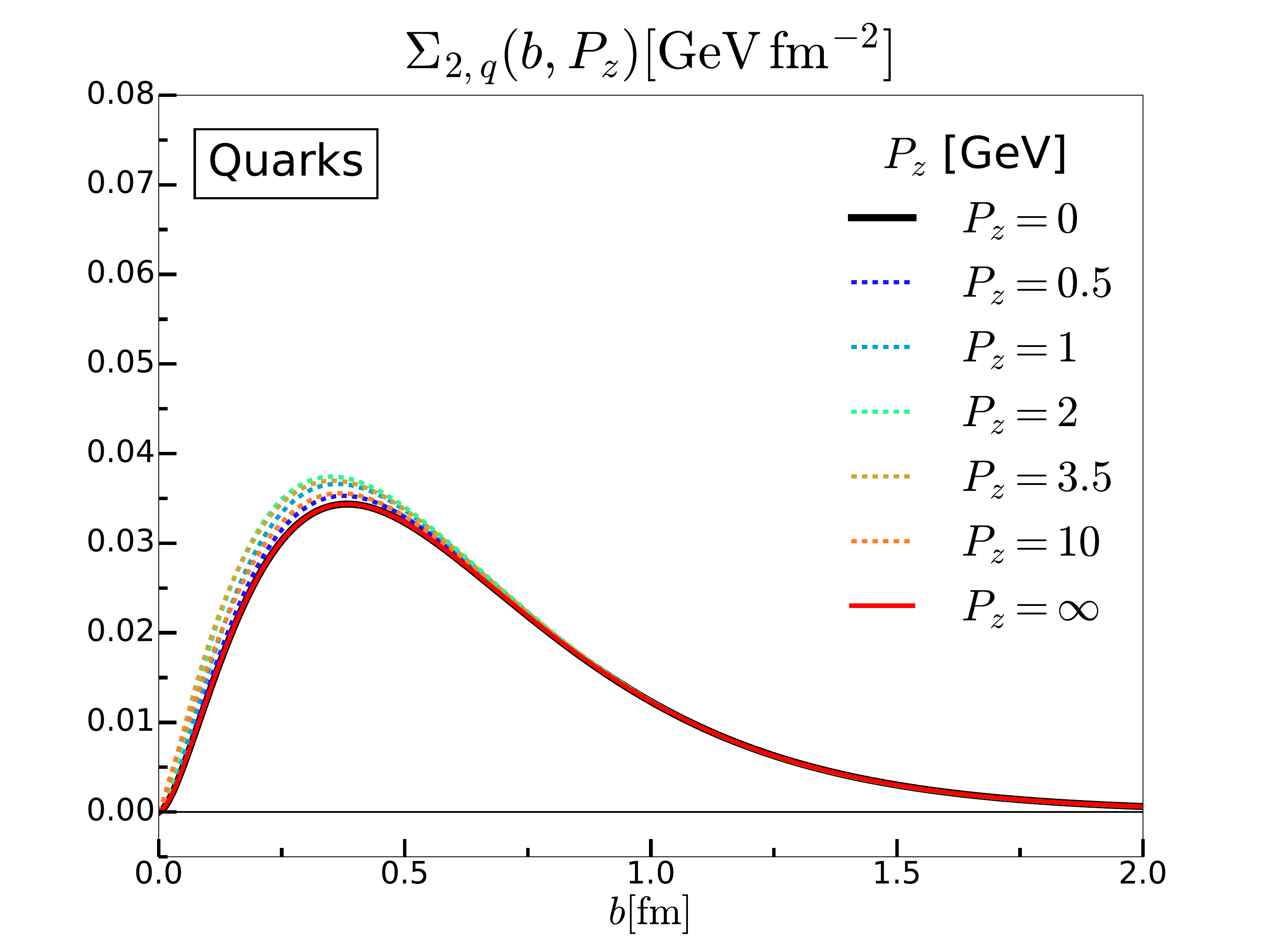}
  \end{minipage}
  \hspace{-0.6cm}
  \begin{minipage}{0.30\textwidth}
    \centering
    \includegraphics[width=\textwidth]{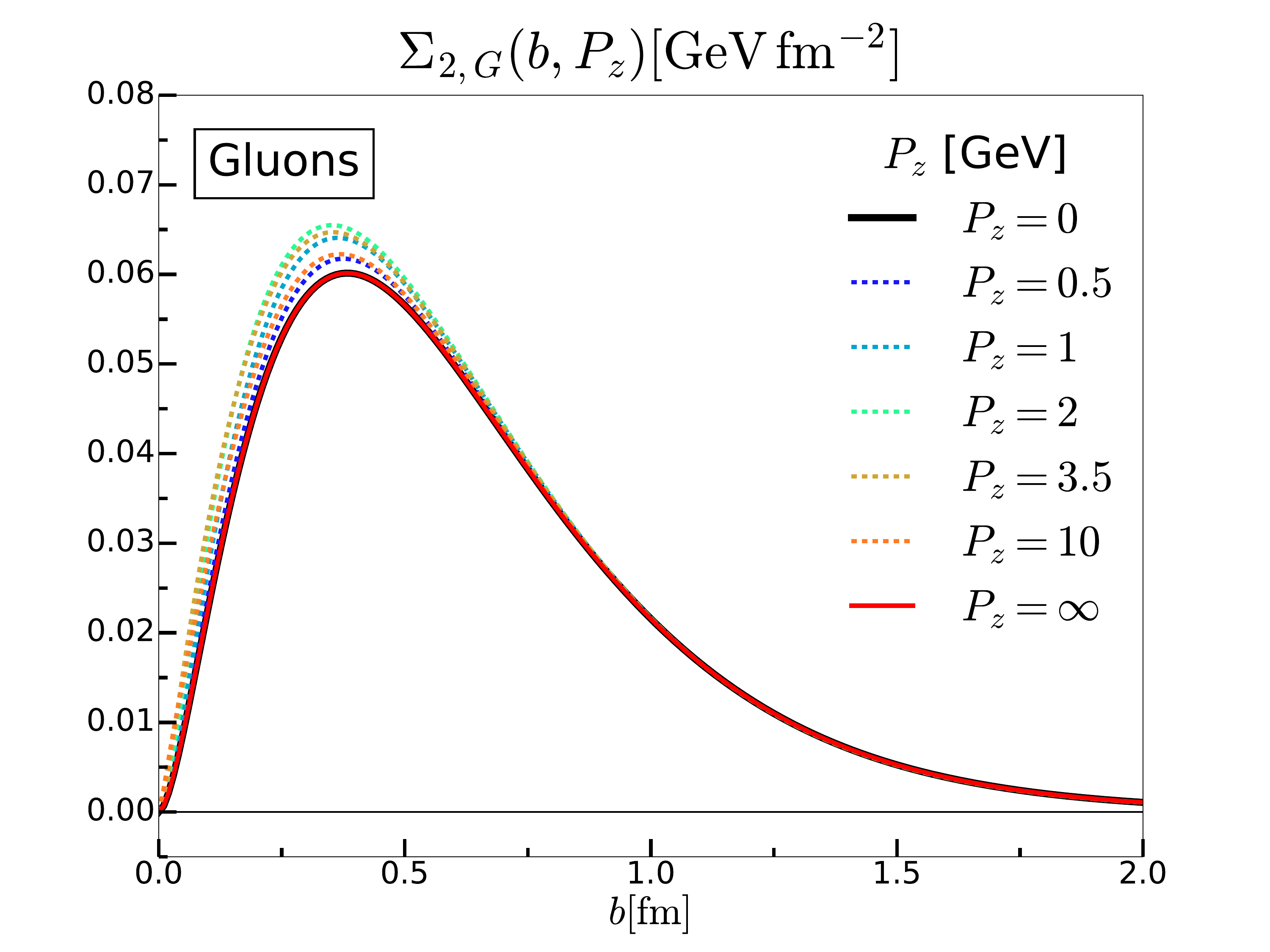}
  \end{minipage}
  \hspace{-0.6cm}
  \begin{minipage}{0.30\textwidth}
    \centering
    \includegraphics[width=\textwidth]{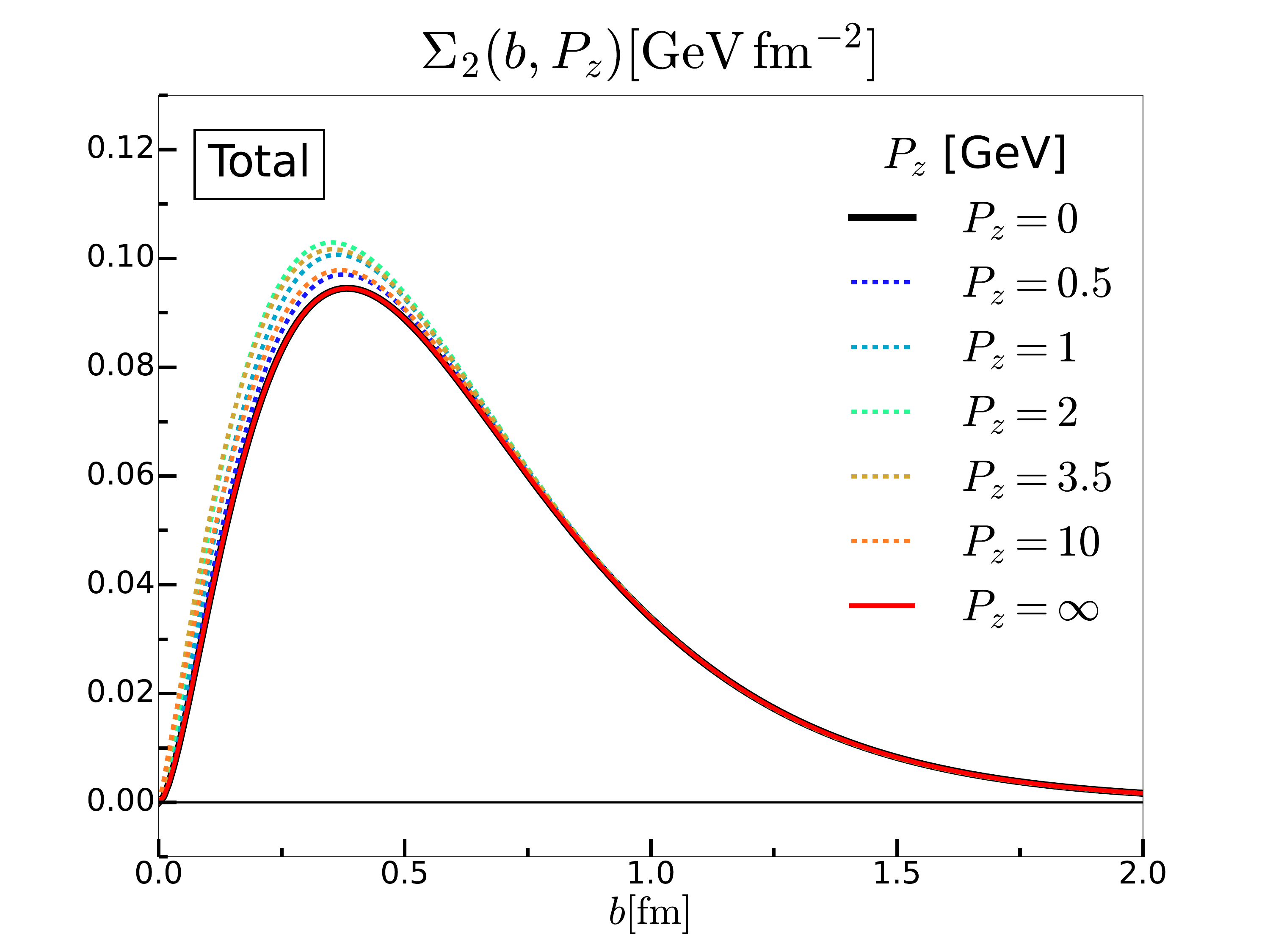}
  \end{minipage}

  \vspace{-0.0cm} 
  
  \begin{minipage}{0.30\textwidth}
    \centering
    \includegraphics[width=\textwidth]{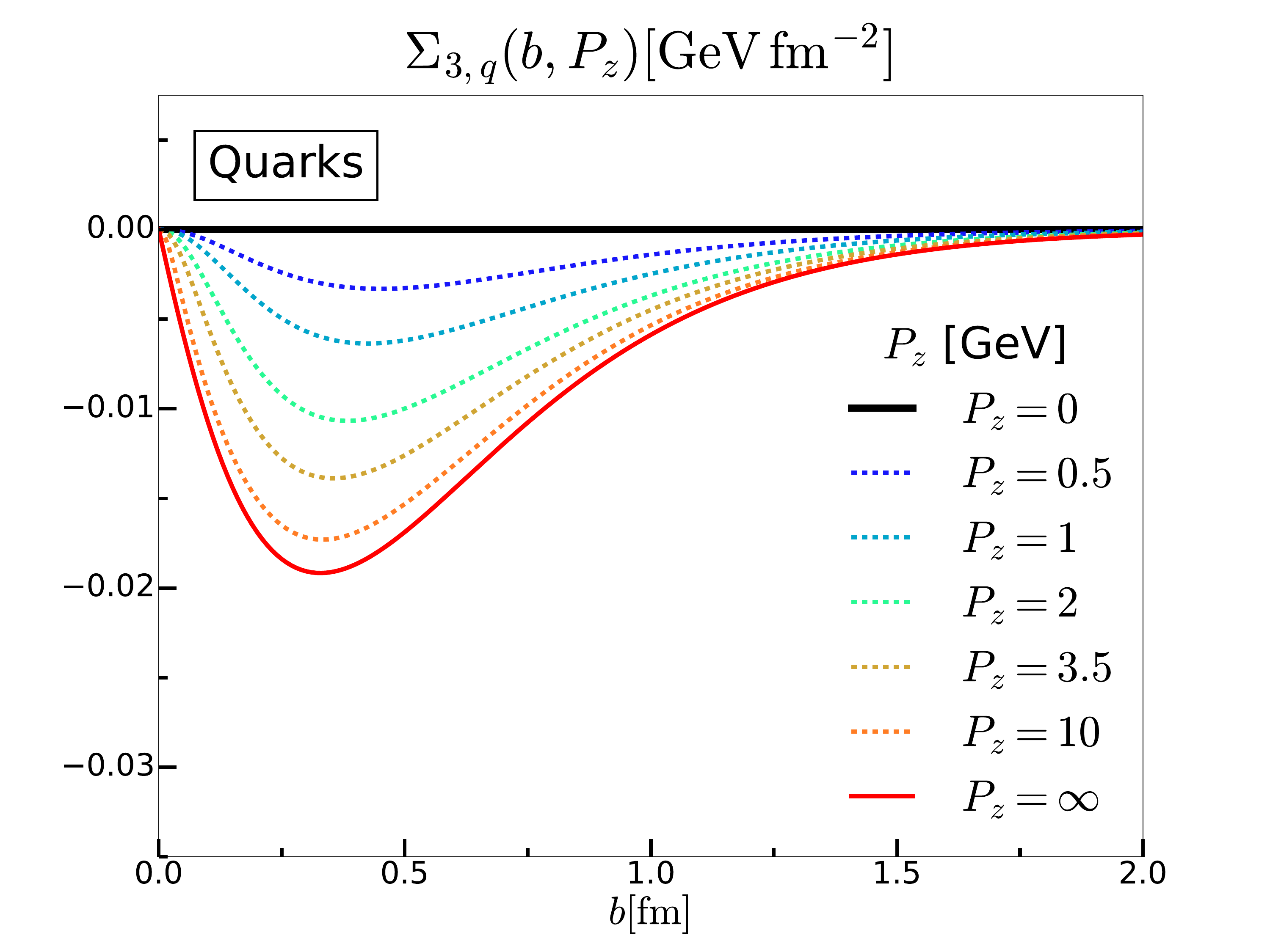}
  \end{minipage}
  \hspace{-0.6cm}
  \begin{minipage}{0.30\textwidth}
    \centering
    \includegraphics[width=\textwidth]{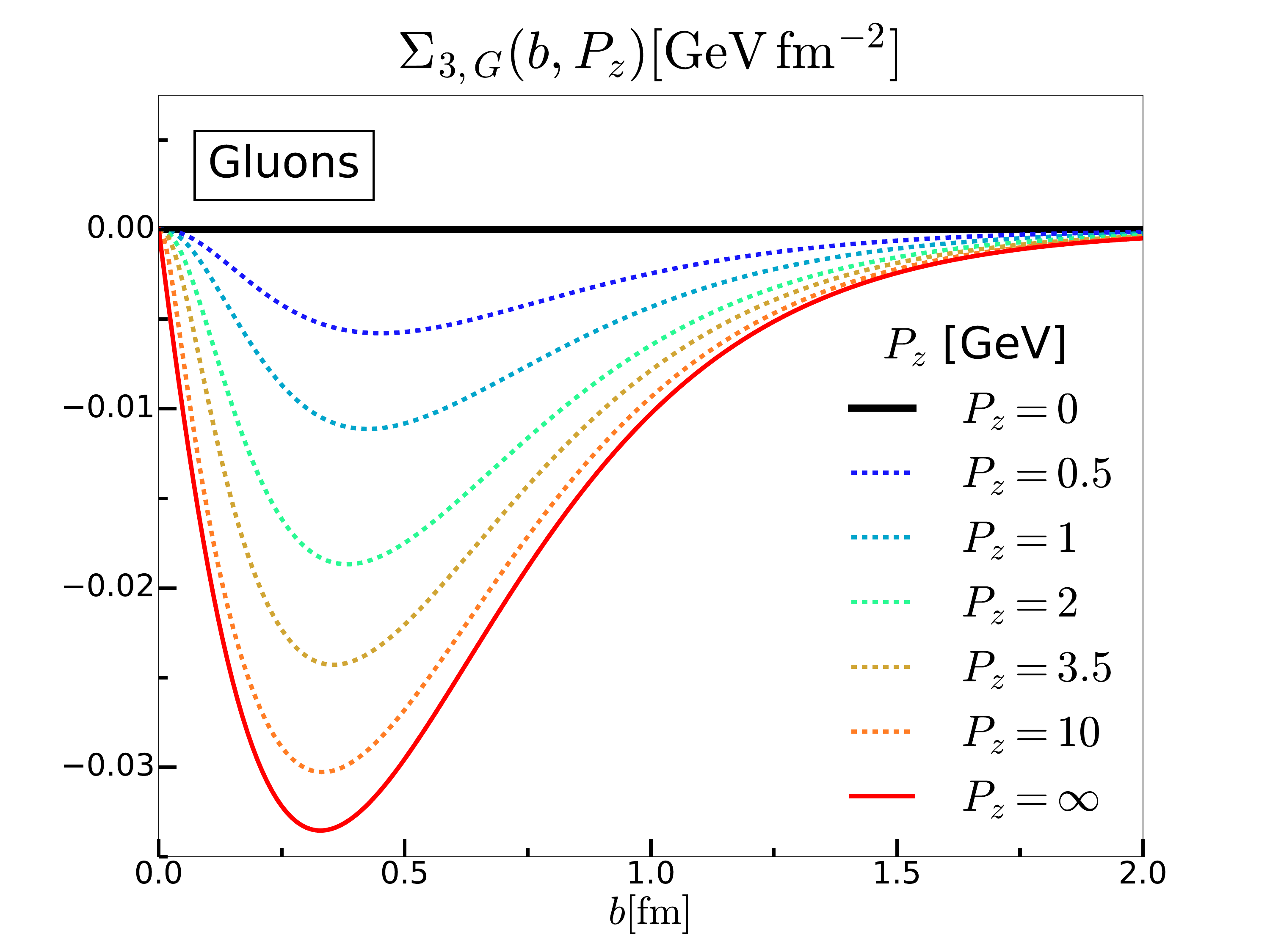}
  \end{minipage}
  \hspace{-0.6cm}
  \begin{minipage}{0.30\textwidth}
    \centering
    \includegraphics[width=\textwidth]{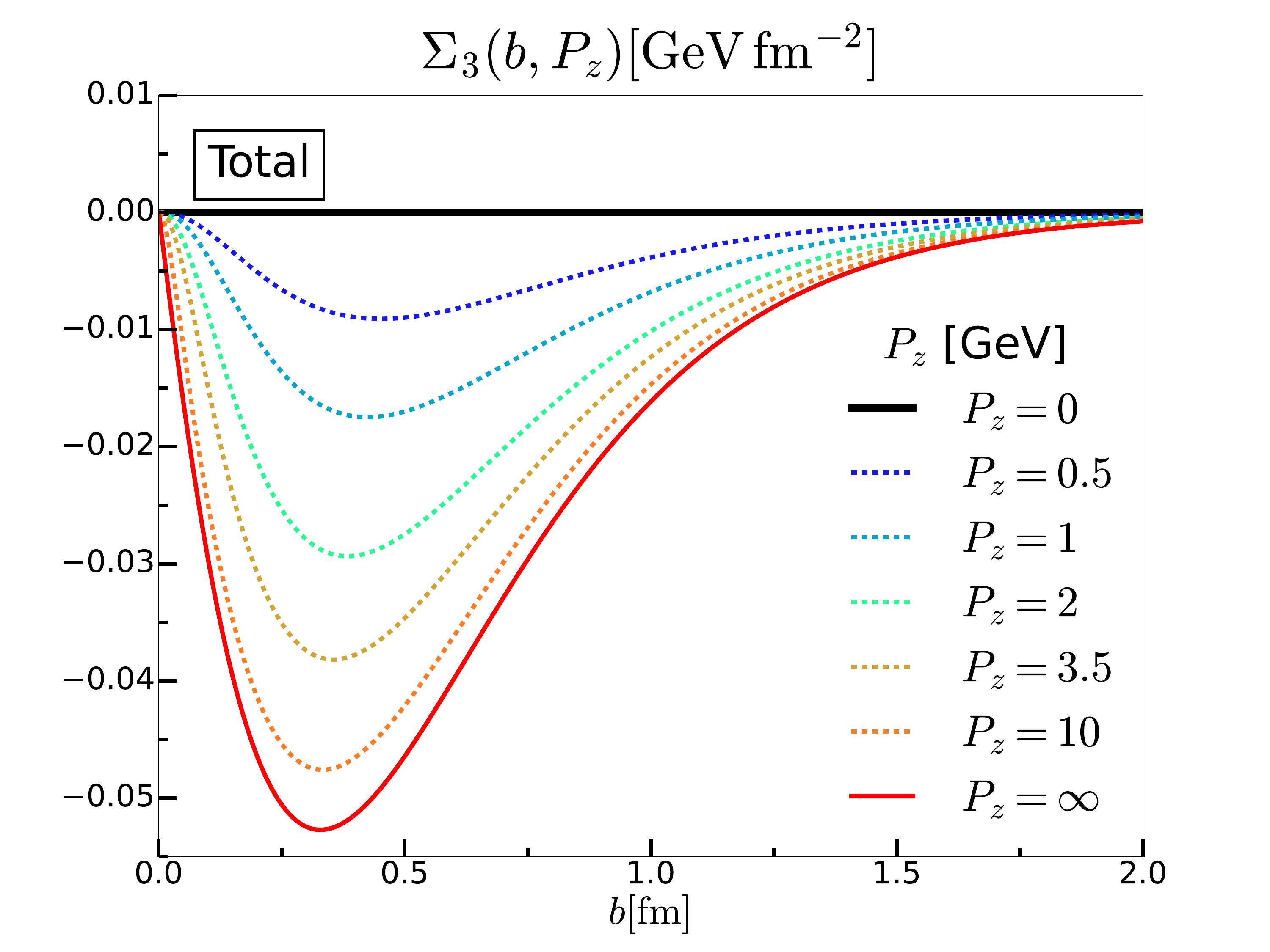}
  \end{minipage}
  
  \caption{EF dipolar (upper row), quadrupolar (middle row), and octupolar (lower row) radial distributions in the transverse plane of 2D anisotropic stress in the nucleon for different values of the nucleon momentum. The first, second, and third columns show the quark, gluon, and total (i.e., quarks + gluons) contributions, respectively. Based on the simple multipole model of Refs.~\cite{Lorce:2018egm,Won:2025dgc} for the EMT FFs. 
}
  \label{fig:5}
\end{figure*}
In Fig,~\ref{fig:5}, we show the three multipolar radial distributions of anisotropic stress for various values of the nucleon momentum. In contrast to the isotropic case, it is the gluon contribution that dominates in the anisotropic case due to the fact that $|D_{G}(Q^{2})| > |D_{q}(Q^{2})|$ in the multipole model of Refs.~\cite{Lorce:2018egm,Won:2025dgc}.

\begin{figure*}[htbp]
  \centering
  \textbf{Total EF distributions of anisotropic stress for a transversely polarized nucleon}

  \vspace{0.5cm} 

  \includegraphics[width=0.90\textwidth]{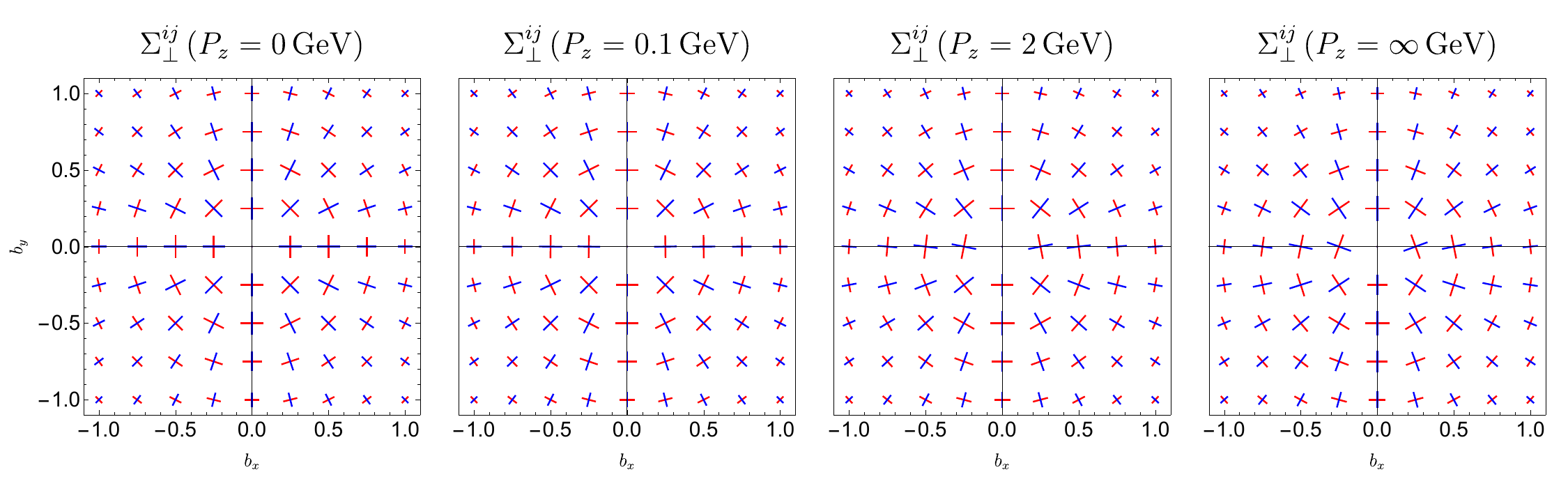}



  \caption{EF distribution of total (i.e., quarks + gluons) transverse anisotropic stress for a nucleon polarized along the $x$-axis and different values of the nucleon momentum. The principal axes of $\Sigma^{ij}_\perp$ at a point $\bm b_\perp$ in the transverse plane are represented by colored segments (blue for positive and red for negative) whose size is proportional to the magnitude of $\Sigma^{ij}_\perp$. Based on the simple multipole model of Refs.~\cite{Lorce:2018egm,Won:2025dgc} for the EMT FFs. 
  }

  \label{fig:6}
\end{figure*} 
In Fig.~\ref{fig:6}, we represent the total EF distribution of anisotropic stress in a nucleon polarized along the $x$-axis as a symmetric traceless rank-2 tensor field in the transverse plane. As with the isotropic stress distribution, increasing $P_z$ induces a dipolar distortion along the negative $y$-axis as a result of the Wigner spin rotation. The orientation of the principal axes may suggest the presence of a convergence point on the negative $y$-axis, but no such point could be identified.

\section{Distributions on the light front\label{sec.4}} 

Although EF distributions are fully relativistic and nicely interpolate between BF and IMF pictures, they cannot in general be interpreted as genuine densities due to relativistic recoil corrections~\cite{Burkardt:2000za}. Genuine densities can, however, be defined in the LF formalism~\cite{Dirac:1949cp,Brodsky:1997de}, where a Galilean subgroup of the Poincar\'e group is highlighted in the transverse plane~\cite{Susskind:1967rg,Kogut:1969xa,Burkardt:2002hr,Miller:2007uy,Miller:2010nz}. 
This formalism has in particular been used to define the LF densities of $T^{++}$~\cite{Burkardt:2002hr,Abidin:2008sb}, where the LF components are denoted as $a^\mu=(a^{+},a^{-},\bm a_{\perp})$ with $a^\pm=(a^0\pm a^3)/\sqrt{2}$, and similarly for other EMT components~\cite{Lorce:2018egm,Freese:2021czn,Freese:2021qtb}.  

In the LF version of the quantum phase-space formalism, the EMT relativistic spatial distributions are defined as~\cite{Lorce:2017wkb,Lorce:2018egm}
\begin{align}
&   \hspace{-0.5cm}
    \mathcal T^{\mu\nu}
    (\bm{b}_{\perp},P^+;\lambda^{\prime},\lambda)
  = \int \frac{d^{2}\Delta_{\perp}}{\left(2\pi\right)^{2}}\,
    e^{-i\bm{\Delta}_{\perp}\cdot\bm{b}_{\perp}}\notag\\
&   \hspace{1.8cm} \times
    \left.
    \frac{_\text{LF}\!\mel{p^{\prime},\lambda^{\prime}}{T^{\mu\nu}(0)}{p,\lambda}_\text{LF}}{2P^+}
    \right|_{\mathrm{DYF}},
\label{EMTdistribution_LF}
\end{align}
where the LF momentum states with definite LF helicities are normalized according to
$_\text{LF}\!\braket{p^{\prime},\lambda^{\prime}}{p,\lambda}_\text{LF}=
2P^{+}\left(2\pi\right)^{3}\delta(p'^+-p^+)\,\delta^{(2)}
(\bm{p}^{\prime}_\perp-\bm{p}_\perp)\,\delta_{\lambda^{\prime}\lambda}$. The 2D LF distributions are constructed in the Drell-Yan frame (DYF), characterized by $\Delta^+=0$ and $\bm P_\perp=\bm 0_\perp$. These distributions do not depend on the LF time $x^+$ because the LF energy transfer $\Delta^-=(\bm \Delta_\perp\cdot\bm P_\perp-\Delta^+ P^-)/P^+$ vanishes in the DYF~\cite{Lorce:2017wkb}. Since the scalar LF distributions $\mathcal T^{\pm\pm}, \mathcal T^{\pm\mp}$ have already been discussed in our previous work~\cite{Won:2025dgc}, we will consider here only the vector and tensor LF distributions.

In the DYF, we find that
\begin{align}
    &_\text{LF}\!\mel{p^{\prime},\lambda^{\prime}}{T^{\pm i}(0)}{p,\lambda}_\text{LF}\big|_\text{DYF}=-2MP^\pm\sigma_{\lambda^{\prime}\lambda}^{3}\notag\\
    &\qquad\times i \epsilon^{ij}_\perp   
     X_{1}^{j}(\phi_{\bm{\Delta}})\,
    \sqrt{\tau}\left[
    J(Q^{2})
  - S(Q^{2})
    \right]
\end{align}
with $P^-=M^2(1+\tau)/(2P^+)$, and similarly for $T^{i\pm}$ with a change of sign for the intrinsic spin contribution. The fact that this expression is identical to the corresponding EF amplitudes
\begin{align}
    &\mel{p^{\prime},s^{\prime}}{T^{\pm i}(0)}{p,s}\big|_\text{EF}=-2MP^\pm\sigma_{s^{\prime}s}^{3}\notag\\
    &\qquad\times i \epsilon^{ij}_\perp   
     X_{1}^{j}(\phi_{\bm{\Delta}})\,
    \sqrt{\tau}\left[
    J(Q^{2})
  - S(Q^{2})
    \right]
\end{align}
can be understood as follows: The EF and DYF conditions being equivalent, EF and DYF amplitudes can only differ by the Melosh spin rotation $\mathcal M_{s\lambda}^{(1/2)}(p)=\bar u( p,s)u_\text{LF}(p,\lambda)/(2M)$ converting canonical polarization to LF helicity~\cite{Melosh:1974cu,Lorce:2011zta}. Just like the Wigner rotation, the Melosh rotation preserves the elements of the third Pauli matrix, i.e.~$\mathcal M_{\lambda^{\prime}s^{\prime}}^{*(1/2)}
    (p')\sigma^3_{s's} \mathcal M_{s\lambda}^{(1/2)}
    (p)=\sigma^3_{\lambda'\lambda}$~\cite{Chen:2022smg}.

For the purely transverse tensor $T^{ij}$, we get
\begin{widetext}
\begin{equation}
 _\text{LF}\!\mel{p^{\prime},\lambda^{\prime}}{T^{ij}(0)}{p,\lambda}_\text{LF}\big|_\text{DYF}
 = 2M^{2} 
    \left[
   \delta_{\lambda^{\prime}\lambda}
  - i\epsilon^{kl}_\perp
    \sigma_{\lambda^{\prime}\lambda}^{k}\,
   X_{1}^{l}(\phi_{\bm{\Delta}})\,
     \sqrt{\tau} \right]\left[\delta^{ij}_\perp F_{0}(Q^{2})+X_{2}^{ij}(\phi_{\bm{\Delta}})\,\tau F_{2}(Q^{2})\right],
    \label{LF_Tij} 
\end{equation}
to be compared with
\begin{equation}
 \mel{p^{\prime},s^{\prime}}{T^{ij}(0)}{p,s}\big|_\text{EF}
= 2M\sqrt{P^2} 
    \left[
   \delta_{s^{\prime}s}\cos\theta
  + i\epsilon^{kl}_\perp
    \sigma_{s^{\prime}s}^{k}\,
   X_{1}^{l}(\phi_{\bm{\Delta}})
     \sin(\theta) \right]\left[\delta^{ij}_\perp F_{0}(Q^{2})+X_{2}^{ij}(\phi_{\bm{\Delta}})\,\tau F_{2}(Q^{2})\right].
    \label{LF_Tij_bis} 
\end{equation}
\end{widetext}
In the IMF, the Wigner rotation reduces to~\cite{Chen:2022smg}
\begin{equation}
    \lim_{P_{z}\to\infty}
    \cos{\theta}
= \frac{1}{\sqrt{1+\tau}}, 
    \quad
    \lim_{P_{z}\to\infty}
    \sin{\theta}
  = - \frac{\sqrt{\tau}}{\sqrt{1+\tau}},
\label{WR_IMF}
\end{equation}
so that the expressions~\eqref{LF_Tij} and~\eqref{LF_Tij_bis} become identical. This is consistent with the fact that canonical polarization and LF helicity become equal in the IMF, i.e.
\begin{equation}
  \lim_{p_z\to\infty}  \mathcal M_{s\lambda}^{(1/2)}(p)=\delta_{s\lambda}.
\end{equation}
This proves once more that LF distributions coincide (up to a normalization factor) with EF distributions in the IMF~\cite{Chen:2022smg,Chen:2023dxp,Won:2025dgc}. When a LF amplitude does not depend on $P^-$, the corresponding LF distribution can be regarded as an actual density~\cite{Lorce:2018egm,Freese:2021czn,Freese:2021mzg,Chen:2022smg}.

\section{Summary \label{sec.5}}

We investigated the relativistic spatial distributions of the energy–momentum tensor for polarized nucleons within the quantum phase–space formalism. 
We focused on the transverse components that had not been addressed in previous studies, and analyzed their relativistic spatial distributions. We showed in particular that when relativistic spatial distributions are decomposed into two-dimensional mutipole contributions, their frame dependence becomes relatively simple. 

While the spatial distributions of transverse momentum and transverse energy flux do not depend on the nucleon momentum, longitudinal boosts induce mixed (i.e.~longitudinal-transverse) stresses whose spatial distributions get distorted by a non-trivial Lorentz factor. However, these distortions disappear in the infinite-momentum frame. In contrast, the spatial distributions of transverse stresses do not mix under longitudinal boosts, but undergo a non-trivial Wigner spin rotation. The latter induces in particular a transverse dipole shift of these distributions when the nucleon is transversely polarized.

Finally, we introduced the corresponding light-front distributions and showed that they coincide (up to a normalization factor) with the infinite-momentum limit of our relativistic spatial distributions. We therefore demonstrated once again that the quantum phase-space formalism allows us to interpolate between the Breit frame and light-front pictures of the nucleon, clarifying in passing the origin of the relativistic distortions.

\section{Acknowledgments \label{sec.6}}

The authors thank Jun-Young Kim for valuable discussions.
The work of H.-Y.W.~is supported by the France Excellence scholarship 
through Campus France funded by the French government 
(Ministère de l’Europe et des Aﬀaires Étrangères), 
Grant No.~141295X.

\bibliography{RED}
\bibliographystyle{apsrev}

\end{document}